\documentclass[preprint,aps,prd, showpacs, amsmath,amssymb,eqsecnum]{revtex4}
\usepackage{graphics}
\usepackage{graphicx}%
\usepackage{wrapfig}
\usepackage{subfigure}
\usepackage{mathptmx}
\usepackage{psfig}
\usepackage{dcolumn}
\usepackage{bm}
\usepackage{tabls}
\newcommand{\beq}{\begin{equation}}
\newcommand{\eeq}{\end{equation}}
\newcommand{\be}{\begin{eqnarray}}
\newcommand{\ee}{\end{eqnarray}}
\long\def\hidestart#1\hideend{}
\begin{document}

\title{Investigations in 1+1 dimensional lattice $\phi^4$ theory}

\author{Asit K. De}
\email{asitk.de@saha.ac.in}

\author{A. Harindranath}
\email{a.harindranath@saha.ac.in}

\author{Jyotirmoy Maiti}
\email{jyotirmoy.maiti@saha.ac.in}

\author{Tilak Sinha}
\email{tilak.sinha@saha.ac.in}

\affiliation{Theory Group, Saha Institute of Nuclear Physics \\
 1/AF Bidhan Nagar, Kolkata 700064, India}

\date{October 6, 2005}

\begin{abstract}
In this work we perform a detailed numerical analysis of (1+1) dimensional
lattice $\phi^4$ theory. We explore the phase diagram of the theory with
two different parameterizations. We find that symmetry breaking occurs only
with a negative mass-squared term in the Hamiltonian. 
The renormalized mass $m_R$ and the field renormalization
constant $Z$ are calculated from both coordinate space and momentum space
propagators in the broken symmetry phase. The critical coupling for the
phase transition and the critical exponents associated with 
$m_R$, $Z$ and the order parameter are extracted using a finite size
scaling analysis of the data for several volumes. The scaling behavior of $Z$
has the interesting consequence that $\langle\phi_R\rangle$ does not scale in
$1+1$ dimensions. We also calculate  the renormalized coupling constant $
\lambda_R$ in the broken symmetry phase. The ratio $ \lambda_R/m_R^2 $ does
not scale and appears to reach a value independent of the bare parameters
in the critical region in the infinite volume limit. 

\end{abstract}


\maketitle


\section{Introduction}
Over the years the $1+1$ dimensional $\phi^4$ theories have been used
and investigated for many
purposes, including theoretical and algorithmic developments in novel
nonperturbative approaches. 
There is a large body of work that deals with the
theory in the continuum starting from the mid-seventies till now
\cite{cphi}. They involve techniques such as Hartree approximation, Gaussian
effective potential, post Gaussian approximations, random phase approximation
and discrete and continuum light front Hamiltonian. 
These studies have been done with a positive bare mass-squared ($m^2>0$) and
diverging contribution to the mass at the lowest non-trivial order of
the coupling ($O(\lambda)$) arising from normal ordering was cancelled
by a counter term, effectively dropping the divergent piece. 
A phase transition to broken symmetry phase was found at strong
quartic coupling. Critical value for $\lambda/m^2$ has been
found to lie somewhere between 30 and 60.

Lattice regularization is naturally suited to determine the
phase diagram of a quantum field theory. There has been an attempt on
the lattice \cite{lw} to extract the critical value of $\lambda/m^2$.
This calculation was performed with a negative bare mass-squared term
in the lattice action resulting
in the broken symmetry phase at small coupling. The
negative mass-squared was converted to a positive mass-squared in the
infinite volume limit by a renormalization performed after the lattice data
had been extracted and a critical value for $\lambda/m^2$ was quoted.

In this paper we investigate the 1+1 dimensional $\phi^4$ theory on
the lattice. To the best of our knowledge, there does not exist any
detailed study of the critical region of this theory using the
nonperturbative numerical program of quantum field theories on the
lattice. Our aim is to explicitly determine the scaling behavior of
the renormalized mass, the renormalized coupling and the field
renormalization constant including their amplitudes. We also want to
investigate the topological sector of this theory in the broken
phase. In a companion work \cite{tcphi} we have calculated the
topological charge using the same
nonperturbative techniques and have shown its relation to the
renormalized parameters in the quantum theory.   
   
In this theory the quartic coupling has the dimension of
mass-squared and a `physical' (relevant in the continuum)
quantity to calculate is the dimensionless ratio of the
renormalized parameters $\lambda_R/m_R^2$. We 
determine the phase diagram in the two dimensional bare
parameter space which agrees with the phase diagram of \cite{lw}. 
We have not found a phase transition from the symmetric phase to the
broken symmetry phase with a positive mass-squared term in the
action. The symmetry breaking occurs in our lattice theory only with a
negative mass-squared term. 
We perform a detailed study of the scaling region of the broken
symmetry phase and determine the ratio
$\lambda_R/m_R^2$ which appears to be constant in
the scaling region irrespective of the
bare lattice parameters. The
vacuum expectation value of the renormalized field
$\langle\phi_R\rangle$ also seems to be constant irrespective of the
parameters in the
scaling region of the $1+1$ dimensional theory. We have determined
these ratios using numerical lattice techniques on a variety of
lattice sizes. We have estimates for their infinite
volume values.

For a nonperturbative approach like the lattice, the notion of perturbative
renormalizability is to be replaced by existence of critical manifolds and
universality classes. The $\phi^4$ theories are generally believed to be in
the same universality class as the Ising model. This notion originally came
from the Renormalization Group and in two Euclidean dimensions is consistent
with a conjecture based on conformal field theory \cite{az}. In our
investigation we determine the critical exponents of $\langle\phi\rangle$ and
$m_R$ independently in
$\phi^4$ theory and find them to be the same as the Ising
values. Another important ingredient relevant in our analysis is the 
field renormalization constant $Z$ which appears in the two point
correlation function. The critical exponent of $Z$ emerging from our FSS
analysis in $\phi^4$ theory is found to be consistent with the
exponent of susceptibility in the Ising model.

Although it is not
the ultimate goal of our work, our results provide an independent
confirmation of the universality of Ising model and $\phi^4$ theory
(with negative mass-squared) in 1+1 dimensions  using numerical
techniques of lattice field theory.
However, in an actual calculation,
always done in a bare theory with an ultraviolet cut-off like
the lattice (which also has a finite size), there are many important
issues still to be resolved, for
example, the onset of the scaling region, effects of the finite size,
possible scaling violation etc. The above needs to be done in each
theory for a complete understanding of the process of the continuum
limit.

In order to determine the ratio $\lambda_R/m_R^2$, we need to know the
field renormalization constant $Z$ and the renormalized mass $m_R$
which can be defined
and determined in two ways: 1) the exponential fall-off in
Euclidean time of the zero-spatial-momentum bare lattice propagators in the
coordinate space, and 2) the behavior of the momentum space propagators
for small four-momenta. We expect respective critical exponents corresponding 
to the renormalized mass and the field renormalization constant to agree
for the two methods and we verify this in the current paper. In this
connection, we wish to point out that we have made use of finite size
scaling for accurate determination of the critical point and
verification and determination of 
the critical exponents. For recent
calculations in 3+1 dimensional Ising model, see \cite{bdwnw}.  

Cluster algorithms, known for beating critical slowing down in Ising
models, are not directly applicable to the $\phi^4$ theories. However, owing
to a development by Wolff \cite{wolff2} using embedded Ising variables
\cite{bt} we have been able to use cluster algorithms in conjunction with the
usual Metropolis Monte Carlo.  

The plan of the paper is as follows. In Sec. II we define the two
parameterizations of the $\phi^4$ theory on the lattice followed by section
III where we discuss the use of embedded Ising variables in $\phi^4$ theory
for use of the cluster algorithms. We show the phase structure of the
lattice theory in section IV. In section V we present the   
calculation of the connected scalar propagator in coordinate space and
momentum space and then in section VI we extract the critical exponents 
for mass $m_R$ and
field renormalization constant $Z$ and determine the critical coupling 
using finite size scaling analysis. 
In Sec. VII the renormalized coupling constant $\lambda_R$ and the
quantities  $\lambda_R/m_R^2$ and $\phi_R$ are discussed.
Finally in Sec. VIII we conclude with a summary of our results. 
  
\section{$\phi^4$ theory on lattice}
In this section we present the lattice action of (1+1) dimensional lattice
$\phi^4$ theory in two different parameterizations.

\subsection{Parameterization as in the Continuum}

We start with the Lagrangian density in Minkowski space (in usual notations)
\be
{\cal L} = \frac{1}{2} \partial_\mu \phi \partial^\mu \phi - \frac{1}{2}
m^{2} \phi^2 - \frac{\lambda}{4!} \phi^4
\ee
which leads to the Lagrangian density in Euclidean space
\be
{\cal L}_E = \frac{1}{2} \partial_\mu \phi \partial_\mu \phi +
\frac{1}{2}
m^2 \phi^2 + \frac{\lambda}{4!} \phi^4 .
\ee
                                                                              
\noindent Note that in one space and one time dimensions, the scalar field
$\phi$ is dimensionless and the quartic coupling
$\lambda$ has dimension of mass$^{2}$. 
                                                                               
The Euclidean action is
\be
S_E = \int d^2x {\cal L}_E.
\ee
Next we put the system on a lattice of spacing $a$ with
\be
\int d^2x = a^2 ~ \sum_x.
\ee
                                                                               
\noindent Because of the periodicity of the lattice sites in a toroidal
lattice, the
surface terms will cancel among themselves (irrespective of the boundary
conditions on fields) enabling us to write
\be
(\partial_\mu \phi)^2 = - \phi \partial_\mu^2 \phi
\ee
and on the lattice
\be \partial_\mu^2 \phi = \frac{1}{a^2} \left [ \phi_{x+ \mu}+ \phi_{x -
\mu}
- 2 \phi_x \right ]~.
\ee
$\phi_{x\pm\mu}$ is the field at the neighboring sites in the 
$\pm\mu$ direction. 
Introducing dimensionless lattice parameters $m_0^2$ and $ \lambda_0$ by
$ m_0^2 = m^2 ~ a^2 $ and $ \lambda_0 = \lambda~a^2 $
 we arrive at the lattice action in two Euclidean dimensions
\be
S = -\sum_x \sum_\mu \phi_x \phi_{x+ \mu} ~+~ (2 + \frac {m_0^2}{2})
~\sum_x ~\phi_x^2 + \frac {\lambda_0}{4!}~\sum_x ~ \phi_x^4~.
\ee
We shall henceforth call this lattice action the
continuum parameterization.

All dimensionful quantities in the following are expressed naturally in the
lattice units, basically meaning that they become dimensionless in the
lattice formulation by getting multiplied by appropriate powers of the
lattice spacing $a$.

\subsection{Another Parameterization}

A different parameterization in terms of field $\Phi$ and parameters
$\kappa$ and $\tilde{\lambda}$, henceforth called
the lattice
parameterization is obtained by setting
\be
\phi = \sqrt{2 \kappa} ~ \Phi, ~~m_0^2 = \frac{1 - 2 \tilde{\lambda}}{\kappa}
-
2 d, ~~\lambda_0 = 6 \frac{\tilde{\lambda}}{\kappa^2}\label{transformation}
\ee
where, $d = 2$ in our case. 
This leads to the lattice action
\be
S^{\prime} = -2 \kappa ~ \sum_x\sum_{\mu} \Phi_x \Phi_{x + \mu} ~+~ \sum_x
\Phi_x^2 ~+~ \tilde{\lambda}~ \sum_x (\Phi_x^2 -1)^2\label{latticeaction}
\ee
where we have ignored an irrelevant constant.

In the limit $ \tilde{\lambda} \rightarrow\infty $, configurations with 
$\Phi^2_x \ne 1$ are suppressed. As a result, field variables assume only 
two values $ \Phi_x \rightarrow \pm1 $ and
$ S^{\prime} $ is reduced to the Ising action $S_{\rm Ising}$ with
\be
S_{\rm Ising} = -2 \kappa~ \sum_x\sum_{\mu} \Phi_x \Phi_{ x+\mu} .
\ee

The lattice action given in Eq. (\ref{latticeaction}) is invariant under the
{\em staggered transformation}
\be
\kappa \rightarrow - \kappa, ~~~~ \Phi_x \rightarrow \Phi_{{\rm st},x}
\ee
where $ \Phi_{{\rm st},x} = (-1)^{x_1+x_2}~ \Phi_x $. 
As a result, if $\kappa_c$ is a critical point, there exists another 
critical point at $ - \kappa_c$. We have three phases: broken phase 
for $\kappa > \kappa_c~\left(\langle\Phi\rangle\ne0,~\langle \Phi_{\rm st}
\rangle=0\right)$, symmetric phase for 
$ -\kappa_c < \kappa < \kappa_c~\left(~\langle\Phi\rangle~=~\langle \Phi_{\rm
st} \rangle=0\right)$ and a {\em staggered broken phase} for 
$\kappa < -\kappa_c~\left(~\langle\Phi\rangle=0,~\langle \Phi_{\rm st}
\rangle\ne0\right)$. 
Note that the staggered broken phase is inaccessible in the
continuum parameterization. 

Wherever possible, we have made use of the  lattice parameterization to 
check the implementation of the our algorithm since it allows a 
cross checking of the critical points. 

\section{Algorithm for updating configurations}
It is well known that most algorithms become extremely inefficient near
criticality (i.e near the continuum limit). This
phenomenon is known as critical slowing down (CSD).
To beat CSD in our $\phi^4$ theory (which has an
{\em embedded} Ising variable as explained below) we have used 
a cluster algorithm (known to beat CSD in Ising-like
systems) to update the embedded Ising variables and combined it 
with the usual Metropolis Monte Carlo algorithm. The variant of the 
cluster algorithm
that we have used is due to Wolff \cite{wolff} and is known as `single cluster
algorithm'.

To see what an embedded Ising variable is and how it is made use of, note
that the part of the $\phi^4$ action that responds to a change of sign is
\begin{eqnarray}
S^I[\phi]&=& -\sum_{x,\mu}\phi_x\phi_{x+\mu} \nonumber \\
&=&-\sum_{x,\mu}|\phi_x||\phi_{x+\mu}| \sigma_x\sigma_{x+\mu}
\nonumber \\
&=&-\sum_{x,\mu}J_{x,x+\mu}\sigma_x\sigma_{x+\mu} 
\label{isaction}
\end{eqnarray}
where $\sigma_x={\rm sign}(\phi_x)$ is called the embedded Ising variable and
$J_{x,x+\mu}=|\phi_x||\phi_{x+\mu}|$ {\em resembles} a
coupling that depends on both position and direction.

Notwithstanding the resemblance, the above action {\em does not} describe an
inhomogeneous, anisotropic Ising model since the couplings $J$ will vary over
configurations. In general one cannot update different aspects of a degree
of freedom ( for example the modulus and sign of $\phi$) separately and
independently. Nevertheless, updating the Ising sector (sign of $\phi$) 
in $\phi^4$ theory is legitimate owing to a result due to Wolff \cite{wolff2}
that we will call Wolff's theorem.

We describe the theory underlying this procedure, not always easy to find
elsewhere. We start by explaining Wolff's theorem .

Consider a group $G$ of transformations $T$ acting on the configurations $C$
of some system:
\begin{equation}
C \rightarrow TC. 
\end{equation}

Now let us consider the group G as an auxiliary statistical system whose 
(micro)states are the group elements \{T\}. We define the induced
Hamiltonian governing the distribution of the auxiliary system by
$H(TC)$ where $H$ is just the Hamiltonian of the original system now
considered as a function of $T$. 

Let us now define an algorithm W with 
transition probabilities $p(C;T\rightarrow T')$ such that \\
\noindent 1) $\sum_Te^{-H(TC)}p(C;T\rightarrow T') 
= e^{-H(T'C)}$, \\
2) $p(TC;T_1\rightarrow T_2) = p(C;T_1 T\rightarrow T_2 T)$. 
 
Wolff's theorem states that a legitimate algorithm for updating the original
system is \\
$\bullet$ Fix $C=C_1$ and $T=I$ (identity transformation), \\
$\bullet$ Update $T_1 \rightarrow T_2$ using the W algorithm, \\
$\bullet$ Assign $C_2=T_2 C_1$ as the new configuration.

We shall now demonstrate that if Wolff's theorem is applied to $\phi^4$
theory with an appropriate choice of G, it is equivalent to updating the
Ising sector independently. 

Take $G=Z_2^N$ where N is the number of sites. 
Elements $T$ of $Z_2^N$ can be represented by Ising variables. 
$T=\{\sigma_x\}$ with $\sigma_x= +1$ or $-1$ so that 
\beq
TC \equiv T \{ \phi_x\} = \{\sigma_x\phi_x\}.
\eeq

The induced Hamiltonian is
\be
H(TC)&=& -\sum_{x,\mu}[\sigma_x\phi_x][\sigma_{x+\mu}\phi_{x+\mu}] \nonumber \\
&=& -\sum_{x\mu}|\phi_x||\phi_{x+\mu}| \sigma_x s_x
\sigma_{x+\mu} s_{x+\mu}\nonumber \\
&=& -\sum_{x,\mu}|\phi_x||\phi_{x+\mu}| s_x'
s_{x+\mu}'\nonumber \\
\ee 
where $\phi_x=|\phi_x| s_x$ and $s_x'=\sigma_xs_x$. \\

The proof of our proposition follows if we note that the above Hamiltonian is
indeed an inhomogeneous, anisotropic Ising Hamiltonian and there is a
1-1 mapping between the variables $s_x'$ and $\sigma_x$.

We have used Wolff's single cluster variant of the cluster algorithm 
to update the Ising variables.
Since the configuration space for the $\phi's$ is much larger than that for 
the Ising variables, to ensure ergodicity, the algorithm was blended 
with the standard Metropolis algorithm. 
The blending ratio used was 1:1 
i.e every cluster sweep was followed up by a Metropolis sweep.  

We summarize the main steps in the algorithm. We start with some
initial configuration for the $\phi$ fields. We then update the sign
of the $\phi$ fields using Wolff`s single cluster algorithm using the
action (\ref{isaction}): We choose some site (seed) at random and
select a group of $\phi$ fields (cluster) around the seed having the
same sign as the field sitting at the seed.  The probability for
selecting a particular field is governed by the action
(\ref{isaction}). This process is called {\em growing a cluster}.  We
flip the sign of the fields belonging to the cluster (the variables
$\sigma$ in (\ref{isaction})) when the cluster is fully grown.
Finally we execute a Metropolis sweep over the entire lattice updating
the full $\phi$ fields. This completes one updation cycle.

Throughout this paper we have used periodic boundary conditions.
However, in the companion work \cite{tcphi} dealing with topological
charge we have used antiperiodic boundary conditions where cluster
algorithms do not work \cite{hasen}.

\section{Phase Structure}
The phase structure is determined by looking at the order parameter
$\langle \phi \rangle$ which takes a nonzero value in the
spontaneously broken phase.  With the cluster algorithm however, since
the sign of the field of all the members of the cluster are flipped in
every updation cycle the algorithm actually enforces tunneling between
the two degenerate vacua in the broken phase. As a result, as an artifact, the
average of $ \phi$ over configurations, i.e., the expectation value becomes 
zero. Thus to get the correct nonzero value for the condensate we measure
$\langle |\phi| \rangle$ where $\phi = \frac{1}{\rm Volume}\sum\limits_{\rm
sites}\phi_x$. To understand the mod let us consider a local order
parameter $\langle \phi_x \rangle$. Since the configurations will be
selected at random dominantly from the neighborhood of either vacua
in the broken phase, $\langle \phi_x \rangle$ will vanish when
averaged over configurations thus wiping out the signature of a broken
phase. If one uses $\langle |\phi_x| \rangle$ as the order
parameter then in the broken phase it correctly projects itself onto
one of the vacua yielding the appropriate non-zero value. The use of
this mod, unfortunately, destroys the signal in the symmetric phase
completely by wiping out the significant fluctuations in sign. However
if we choose to use $\langle |\frac{1}{\rm Volume}\sum\limits_{\rm
sites}\phi_x| \rangle$, it correctly captures the broken phase as well as
the symmetric phase.  While the sign fluctuation over configurations are
still masked, the fluctuations over sites survive producing $\langle |\phi|
\rangle = 0$ correctly in the symmetric phase. 

\begin{figure}[tp]
\centering 
\includegraphics[width=3in,clip]{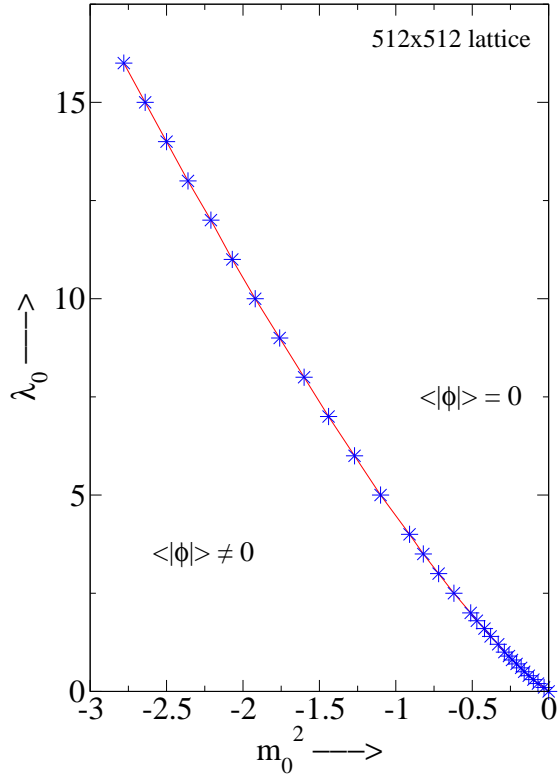}
\caption{Phase diagram for continuum parameterization}
\label{pd_cont}
\end{figure}

The phase diagram for the continuum parameterization obtained for a $512^2$
lattice is presented in Fig. \ref{pd_cont}. Classically, spontaneous symmetry
breaking (SSB) occurs for negative $m_0^2$. For small negative $m_0^2$, as two
minima are shallow and very close to each other, quantum fluctuations
can restore the symmetry. So, larger negative values of $m_0^2$ are required
for SSB to take place. Consequently, the phase transition line is found in the
negative $m_0^2$ semiplane. Our phase diagram agrees qualitatively with
that obtained for much smaller lattices in \cite{ct,aw}. In \cite{lw}
the authors have extrapolated
their results to infinite volume. We find that our $512^2$ lattice results are
as good as the infinite volume result in \cite{lw}.
 
\begin{figure}[tp]
\centering  
\subfigure{
\includegraphics[width=3in,clip]{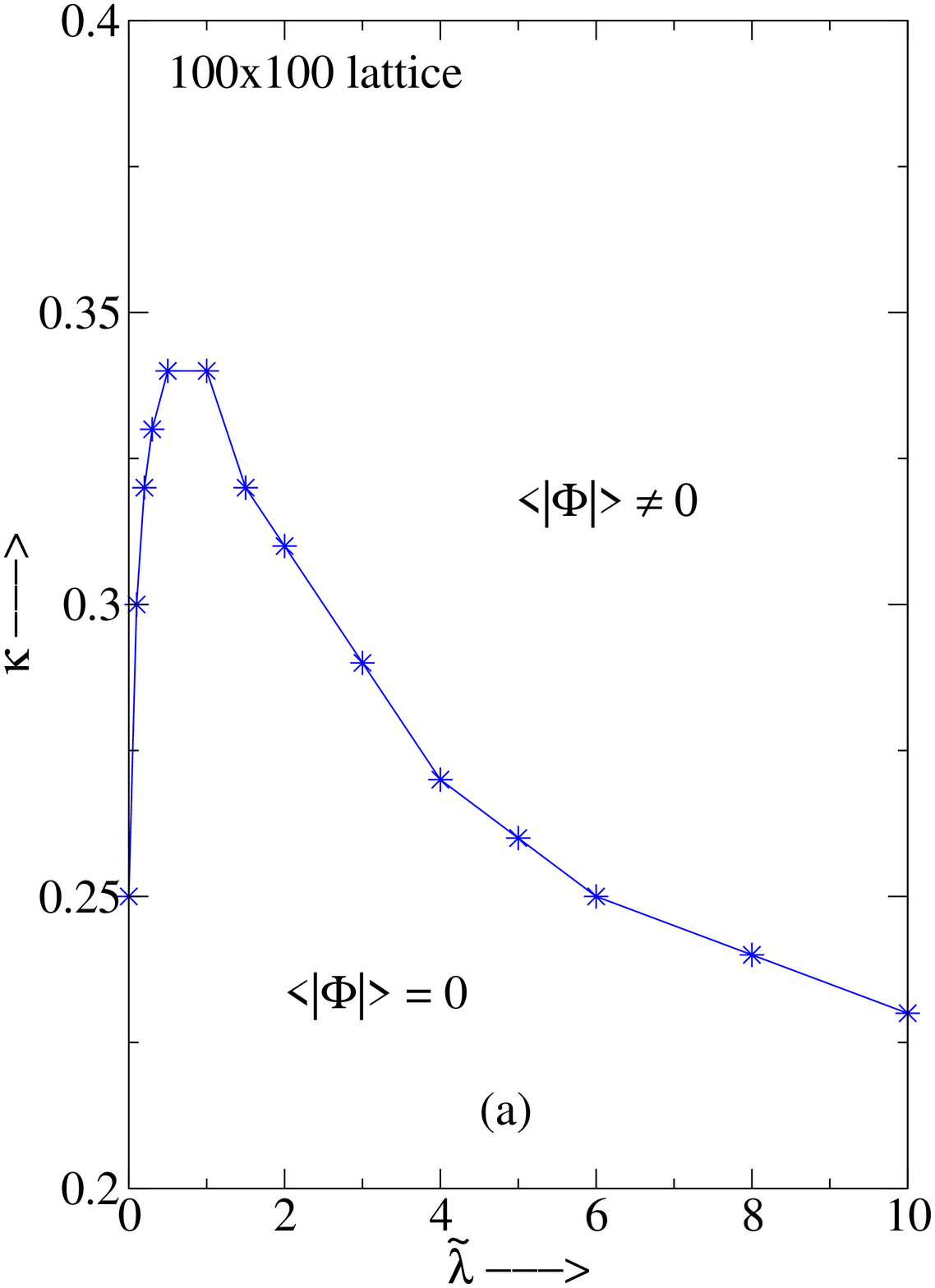}}
\subfigure{       
\includegraphics[width=3in,clip]{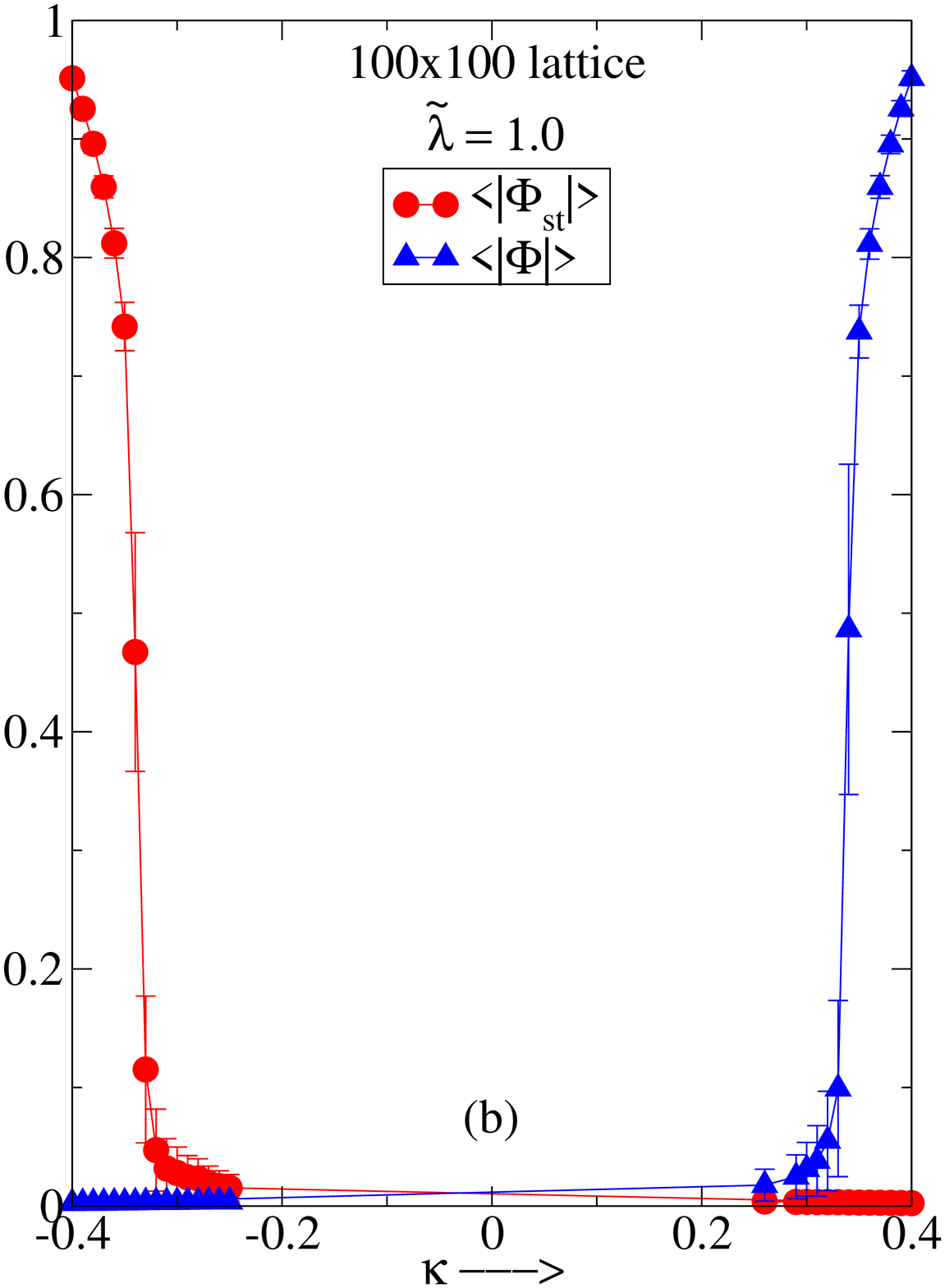}}
\caption{(a) Phase diagram for lattice parameterization. (b) Manifestation of
  the staggered symmetry of lattice parameterization of the action. }
\label{pd-latt}
\end{figure}

In lattice parameterization, phase diagram obtained for a $100^2$
lattice is presented in Fig. \ref{pd-latt}(a). We have restricted ourselves
only to $\kappa \ge 0$ region. The symmetry of the phase diagram for 
$\kappa < 0$ and $\kappa > 0$ is evident from the 
behavior of $ \langle \Phi \rangle $ and $\langle \Phi_{\rm st} \rangle $ as a
function of $\kappa$, shown in Fig. \ref{pd-latt}(b).
  
\begin{figure}
\begin{minipage}[t]{8cm}
\includegraphics[width=1\textwidth]{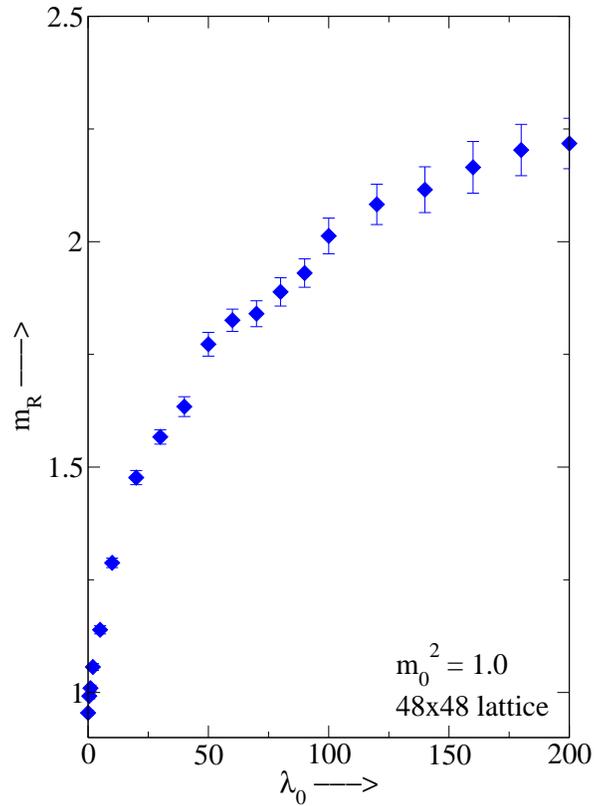}
\caption{Mass gap for $+$ ve $m_0^2$.}
\label{msq_positive}
\end{minipage}
\end{figure}

As mentioned in the Introduction,
in the continuum version of $1+1$ dimensional $ \phi^4$ theory with a positive 
mass-squared  term in the Hamiltonian, there have
been many attempts to calculate critical couplings for phase transition
from the symmetric phase to the broken phase \cite{cphi}.
We have  investigated the phase diagram of the lattice theory in the region
of positive mass-squared and have been unable to detect a phase transition
in this region of the parameter space. In Fig. \ref{msq_positive} we show
the measurement of the mass gap $m_R$ extracted from coordinate space
propagator in positive mass-squared region. We find that
the mass gap $m_R$ monotonically increases with the coupling. 

An alternative way to probe the same region of the parameter space of the
continuum parameterization is to perform simulations with the lattice
parameterization of the action. 
From the phase diagram for the latter presented in Fig. \ref{pd-latt}(a), we 
reconfirm the 
absence of phase transition for the lattice theory in the positive 
mass-squared region in the continuum parameterization since this whole
region can be mapped onto the
symmetric phase in the lattice parameterization using the transformation Eq.
(\ref{transformation}).

\section{Calculation of Propagator}
We have made use of two-point connected correlation function 
to calculate the fundamental boson mass and field renormalization constant. 
We have carried out our simulation both in coordinate and momentum space. 

\subsection{Coordinate space}
In coordinate space, 2-point connected correlation function
$G_c\left(x,x_0\right)$ is given by 
\be
G_c(x,x_0) &=& 
\left<\phi\left(\vec{x},t\right)\phi\left(\vec{x_0},t_0\right)\right> -
\left<\phi\left(\vec{x},t\right)\right>\left<\phi\left(\vec{x_0},t_0\right)
\right>\nonumber\\
 &=& \left<\phi\left(\vec{x},t\right)\phi\left(\vec{x_0},t_0\right)\right> -
\left<\phi\right>^2. \label{gcx}
\ee
For the derivation of the last equation translational invariance has been
assumed. As explained before, we actually calculate
$\left<|\phi|\right>$ instead of $\left<\phi\right>$ in Eq. (\ref{gcx})
The notation $\vec{x}$ may be
confusing in 1+1 dimensions; however, it is kept to distinguish between
the spatial and temporal directions.

We have extracted the renormalized scalar mass $m_R$ (pole mass) and the field
renormalization constant $Z$ from
\begin{eqnarray}
G_c\left(t\right) =
\frac{Z}{2m_R}\left[e^{-m_Rt}+e^{-m_R\left(L-t\right)}\right]
+ {\rm{~higher~states}}.
\label{eq:c1}
\end{eqnarray}
where $G_c\left(t\right)$ is the zero spatial momentum projection of the 
2-point connected correlation function. The second exponential term in
the RHS of the above equation is due to the periodicity of the lattice.

\begin{figure}
\begin{minipage}[t]{8cm}  
\includegraphics[width=1\textwidth]{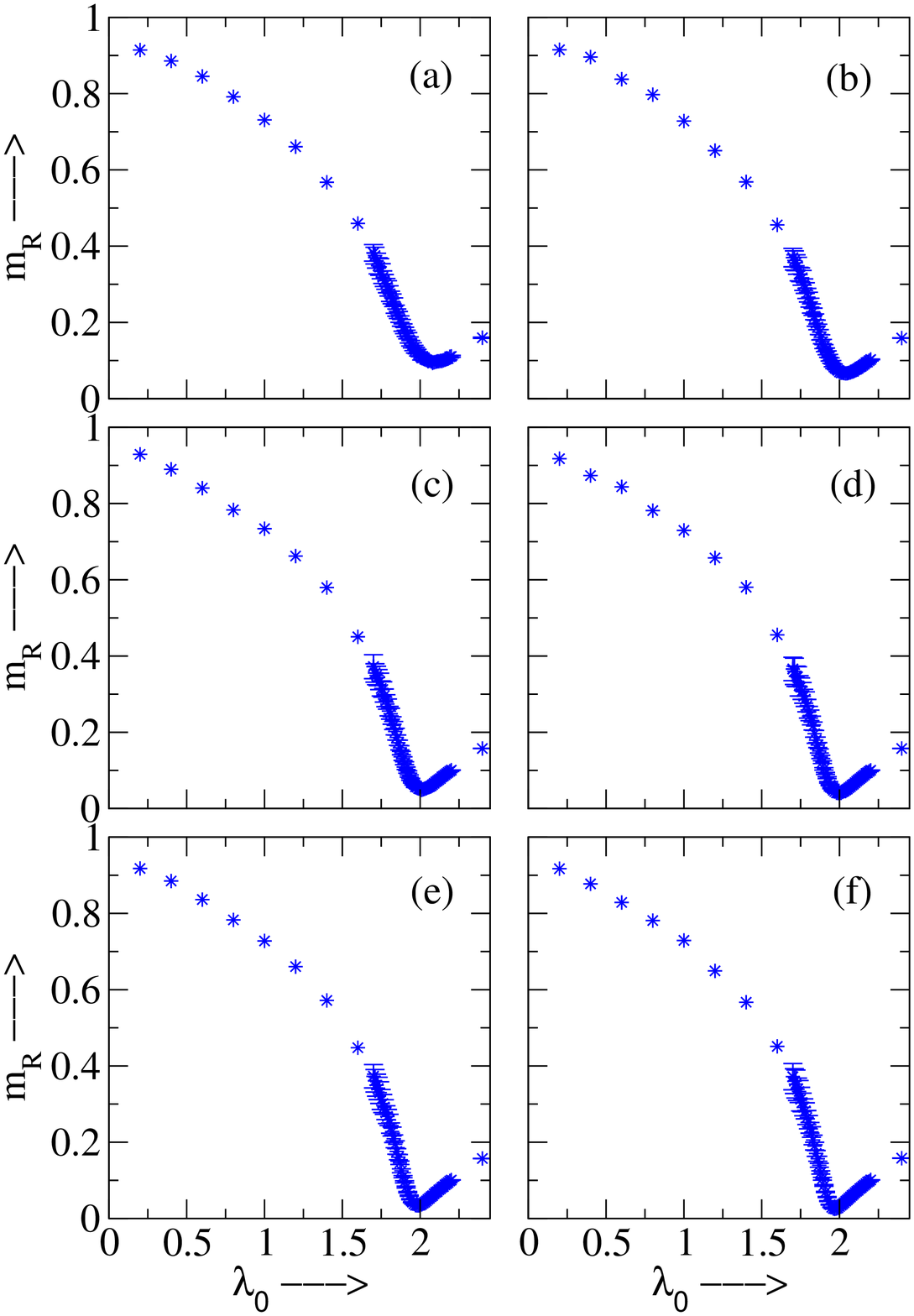}
\caption{$m_R$ from coordinate space propagator for different lattices
(a) $32^2$, (b) $48^2$, (c) $64^2$, (d) $80^2$, (e) $96^2$ and (f) $128^2$
($m_0^2=-0.5$)}
\label{coord_m}
\end{minipage}
\hfill  
\begin{minipage}[t]{8cm} 
\includegraphics[width=1\textwidth]{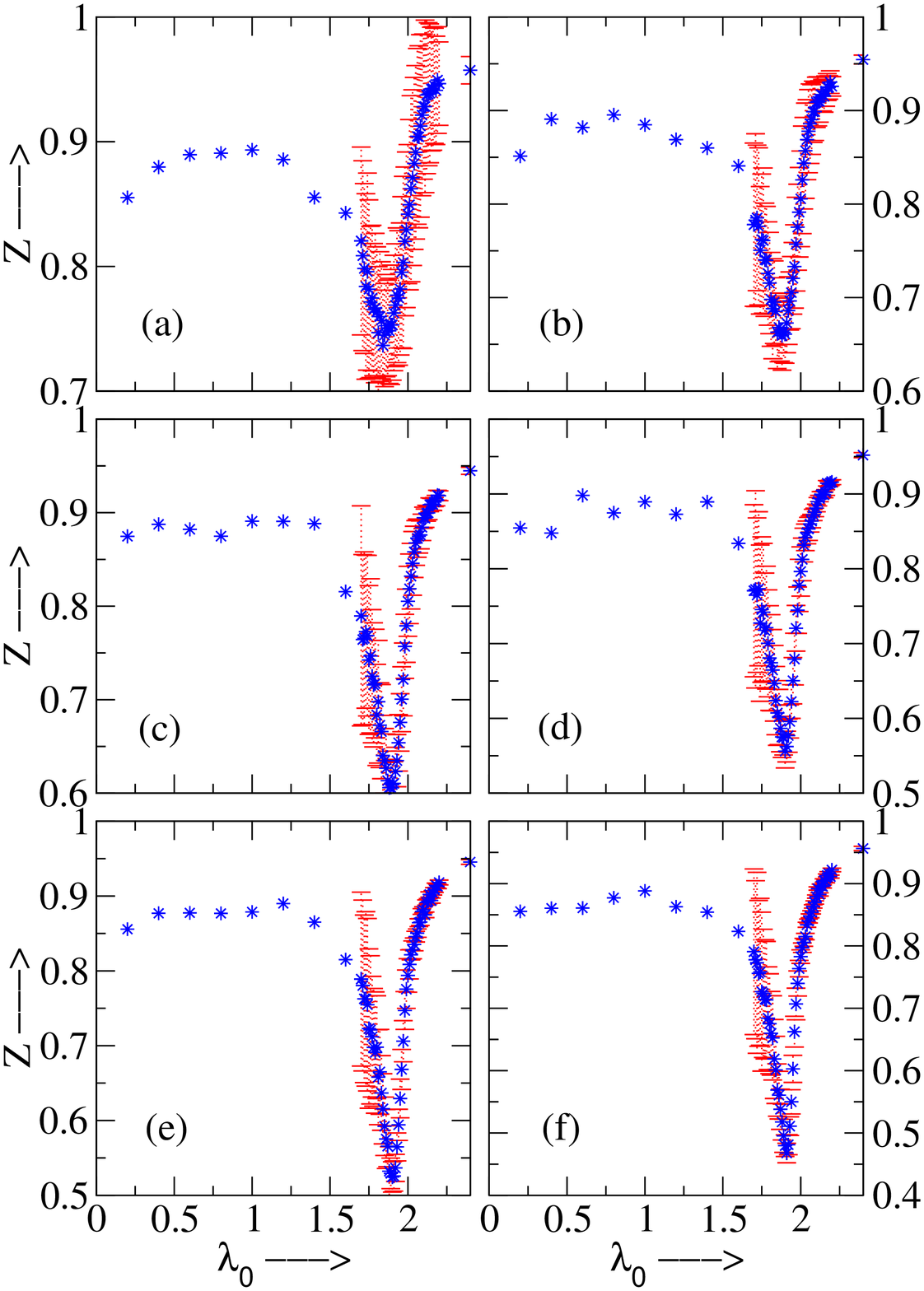}
\caption{$Z$ from coordinate space propagator for different lattices
(a) $32^2$, (b) $48^2$, (c) $64^2$, (d) $80^2$, (e) $96^2$ and (f) $128^2$
($m_0^2=-0.5$)}                                   
\label{coord_z} 
\end{minipage}
\end{figure}  

In this calculation, to ensure thermalization, we
have discarded the first $10^6$ configurations before starting our
measurements. Measurements were carried out on $100$ bins of
$2\times 10^5$ configurations. In each bin, to fulfill the 
requirement that measurements be made on statistically
independent configurations, measurements were performed every 
tenth configuration (hop-length). 

Figs. \ref{coord_m} and \ref{coord_z} show
$m_R$ and $Z$ extracted from the connected propagator in coordinate space for
lattices of six different sizes as function of $\lambda_0$ for fixed 
$m_0^2=-0.5$. The figures clearly demonstrate scaling
of $m_R$ and $Z$ as one moves towards the critical point. The critical point
is given roughly by the dip in each curve. These dips are clearly not at the
same place for $m_R$ and $Z$ at the smaller lattices. A phenomenological
finite size scaling analysis has been done in the next section on these data 
and the data obtained from momentum space propagators to calculate the
critical exponents and the critical coupling in the infinite volume limit.   
Around the critical region in Figs. \ref{coord_m} and \ref{coord_z}, all
the curves have a thick appearance because of the proximity of the many data 
points with the associated errors shown for each point. Outside the scaling
region, a region not of interest to us, the data is relatively sparse and we
also have suppressed the error bars for them.    
Unlike in the Ising model ($\lambda_0\rightarrow\infty$), at finite $\lambda_0$
the $\langle|\phi|\rangle$ takes on large values outside the scaling
region in the broken phase.
In calculating the connected propagator, one
performs a subtraction between the large expectation values of two quantities
measured independently. This enhances the error bars outside the scaling
region in the broken symmetry phase.    

\subsection{Momentum space}
Connected propagator in momentum space is 
\be
G_c(p) = \sum_x e^{ipx}\left[\left<\phi_x\phi_0\right>
  - \left<\phi_x\right>\left<\phi_0\right>\right].
\ee
To improve statistics in numerical simulation, averaging over source
$y$ is performed.  
\be
G\left(p\right) &=&
\frac{1}{V}\sum_{x,y}e^{ip\left(x-y\right)}\left[\left<\phi_x\phi_y\right>
  -\left<\phi_x\right>\left<\phi_y\right>\right]
\nonumber\\&=& 
\left<\frac{1}{V}\sum_{x,y}\phi_x\phi_y\cos p\left(x-y\right)\right>
- \left|\left<\phi\right>\right|^2\delta\left(p\right).
\ee
At small momenta, the momentum space propagator behaves as
\be
G\left(p\right) = \frac{Z^\prime}{m_{R^\prime}^2 + {\hat{p}}^2}
\label{g_p}
\ee
where, ${\hat{p}}^2 =
4\sum\limits_{\mu}\sin^2\left(\frac{p_{\mu}}{2}\right)$ with $\mu =
1,2$ is the dimensionless lattice equivalent of the momentum square in the
continuum.

From the intercept of inverse propagator on the ordinate and slope
at $\hat{p}^2=0$, $m_{R^\prime}$ and $Z^\prime$ can be determined.
However it is only near the critical
coupling that the pole of the propagator is actually near zero and it is here
that $m_{R^\prime}$ approaches the pole mass $m_R$. A similar argument
applies to $Z^\prime$
and $Z$. We have thus calculated the momentum space propagators only near the
critical point.

\begin{figure}[tp]
\centering  
\subfigure{
\includegraphics[width=3in,clip]{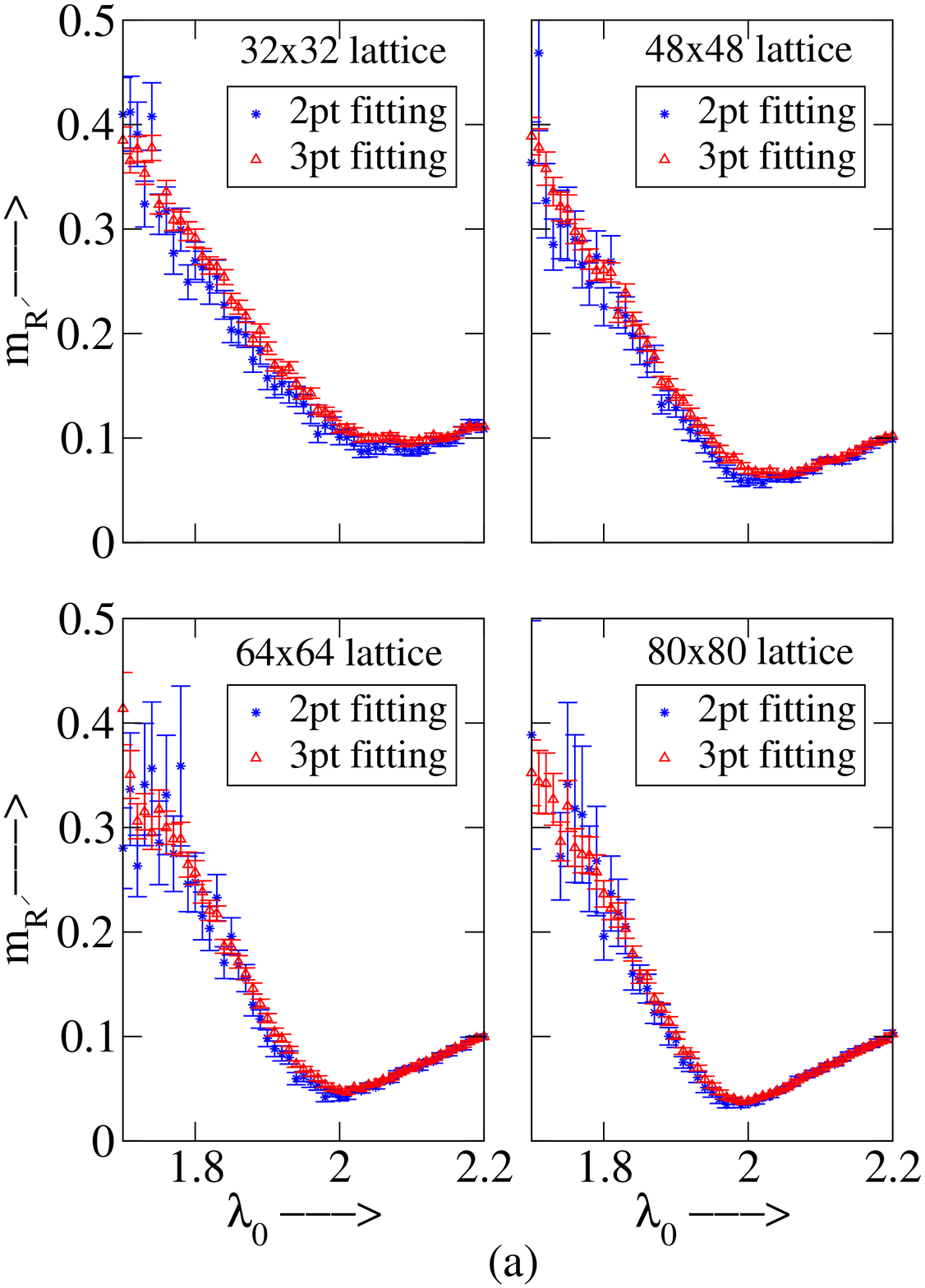}}
\subfigure{       
\includegraphics[width=3in,clip]{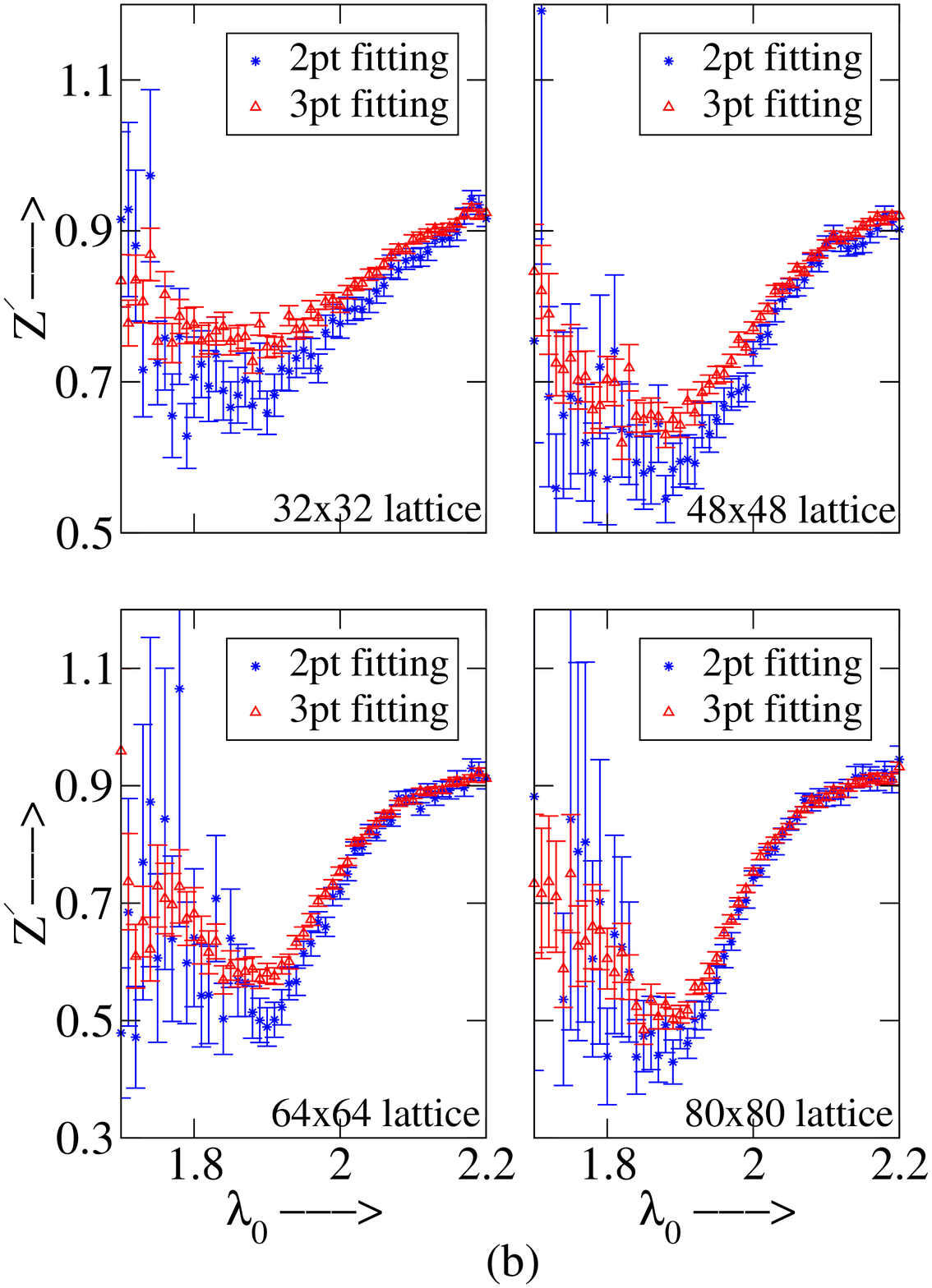}}
\caption{(a) $m_{R^\prime}$ in the critical region for different $L$ from
momentum space propagator, (b) $Z^\prime$ in the critical region for different
$L$ from momentum space both for $m_0^2 = -0.5$}
\label{mom_mz}
\end{figure}   

Since the calculation of momentum space propagators were extremely 
time consuming, we had to restrict ourselves to  smaller lattices.
For thermalization $10^5$ configurations were discarded before starting
the measurements. 
The number and size of bins as well as the hop-length were the same as that
for coordinate space propagators.

As the inverse propagator was found to be nonlinear in the small momenta
region, one could use only the lowest few momentum modes for the determination
of $m_{R^\prime}$ and $Z^\prime$. For this calculation we took the lowest $2$
and $3$ modes  excluding the zero momentum mode because the connected
propagator value for the zero mode is prone to relatively large statistical 
error arising
from the subtraction, in the critical region, between two quantities 
measured independently 
\cite {bdwnw}. In Figs. \ref{mom_mz} (a) and (b), we have presented
$m_{R^\prime}$ and $Z^\prime$ extracted from momentum space propagator with
$m_0^2 = -0.5$ for four different lattices. Results obtained using $2$-point
and $3$-point fitting are found to be quite close to each other. The $3$-point
fitting is more stable and has been used in our study of finite size
scaling in the next section.        

\section{CRITICAL EXPONENTS AND CRITICAL COUPLING FROM FINITE 
SIZE SCALING ANALYSIS }

In this section, we perform a phenomenological Finite Size Scaling (FSS) 
analysis of our data to
extract critical exponents and critical coupling. Let us first briefly
summarize the main aspects \cite{cardy} of this FSS analysis. 
In a finite size system,
there are, in principle, three length scales involved: correlation length $
\xi$, size of the system $L$ and the microscopic length $a$ 
(lattice spacing). FSS assumes that
close to a critical point, the microscopic length $a$ drops out.  
According to FSS \cite{brezin},  for an observable $P_L$ (whose infinite
volume limit displays nonanalyticity at the critical point $\lambda_0^c$), 
calculated in a finite size of linear dimension $L$,
\be
P_{L}(\tau)/P_{\infty}(\tau) = f(L/\xi_{\infty}(\tau)) \label{fssh} 
\ee
where $ \tau = (\lambda_0^c - \lambda_{0})/\lambda_0^c$ and  the 
function $f$ (commonly known as scaling function) is universal in the 
sense that it does not depend on
the type of the lattice, irrelevant operators etc. It does depend on the
observable $P$, the geometry, boundary conditions etc.
For fixed $L$ as $ \tau \rightarrow 0$, strictly there is no phase transition.
Consequently $P_{L}(\tau)$ is not singular at $\lambda_0=\lambda_0^c$. 
Near the critical
point we have, $ \xi_{\infty}(\tau) = A_{\xi} \tau^{-\nu}$ where $ \nu$ is the
critical exponent associated with the correlation length. Suppose, near the
critical point,  $ P_{\infty}(\tau) = A_P
\tau^{-\rho}$. Then from Eq. (\ref{fssh}) 
\be
P_{L}(\tau) = A_P \tau^{-\rho}~ f(A_{\xi}^{-1}L ~\tau^\nu)~.
\ee
Since $ P_{L}(\tau)$ should have smooth behavior as 
$ \tau \rightarrow 0$, a simple ansatz for $f$ may be taken as 
$ f(L~\tau^{\nu}) \sim (L~ \tau^\nu)^{\rho/\nu}$ 
so that $P_{L}(\tau)$
does not blow up as $ \tau \rightarrow 0$. Thus 
$f(x) \sim C x^{\rho/\nu}$ as $x \rightarrow 0$. 

Alternatively, we may write 
\be
P_{L}(\tau) = A_P~A_{\xi}^{-\rho/\nu}~ L^{\rho/\nu}~
g(A_{\xi}^{-1/\nu}~\tau ~L^{1/\nu}),
\ee
where $g$ is another scaling function. Since
$P_{L}(\tau)$ should have no singularity as $ \tau \rightarrow 0$, 
$L$ finite, we have, $g(x) \rightarrow {\rm constant} $  
as $ x \rightarrow 0$. 

\begin{figure}[tp]
\centering  
\subfigure{
\includegraphics[width=3in,clip]{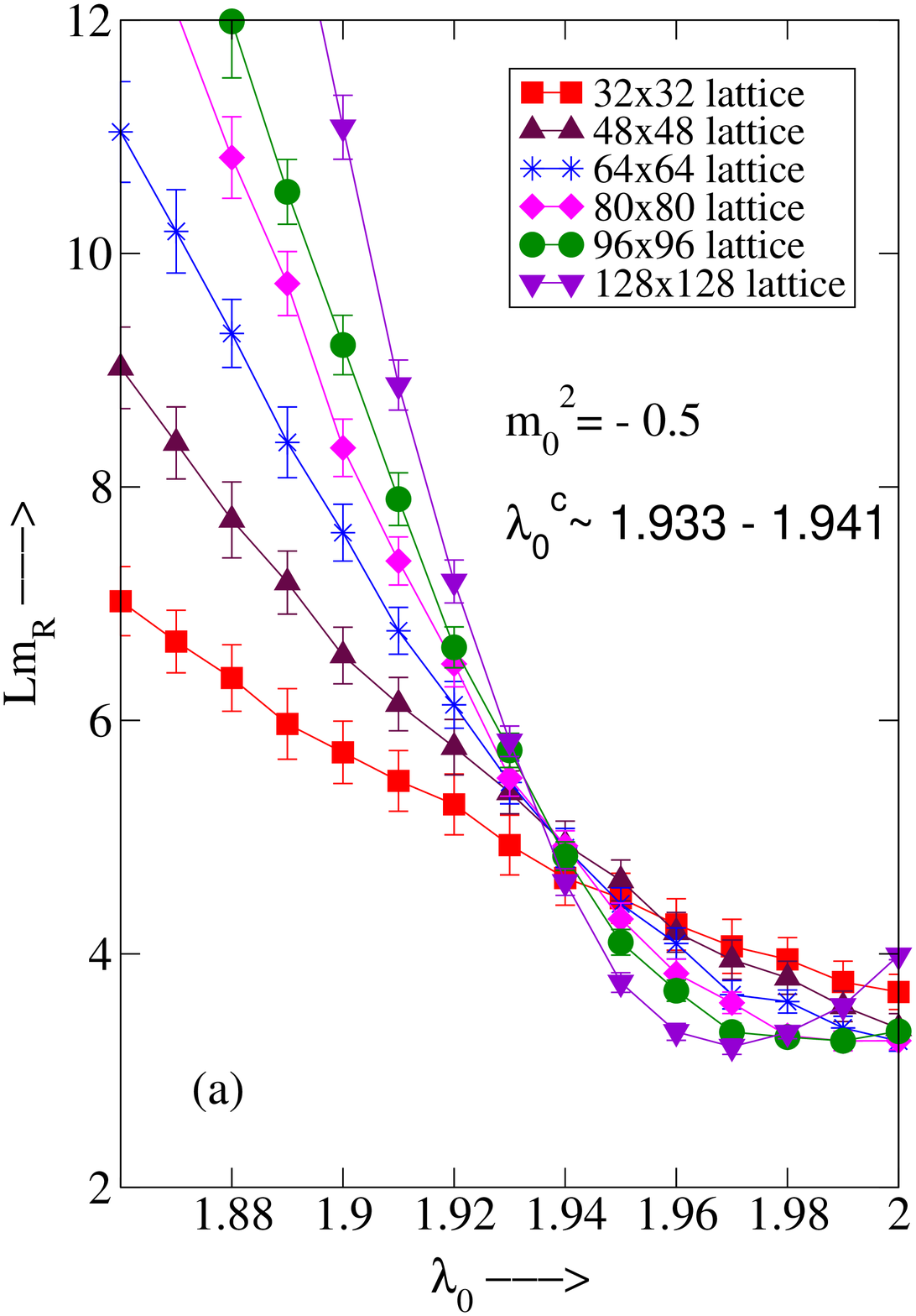}}
\subfigure{       
\includegraphics[width=3in,clip]{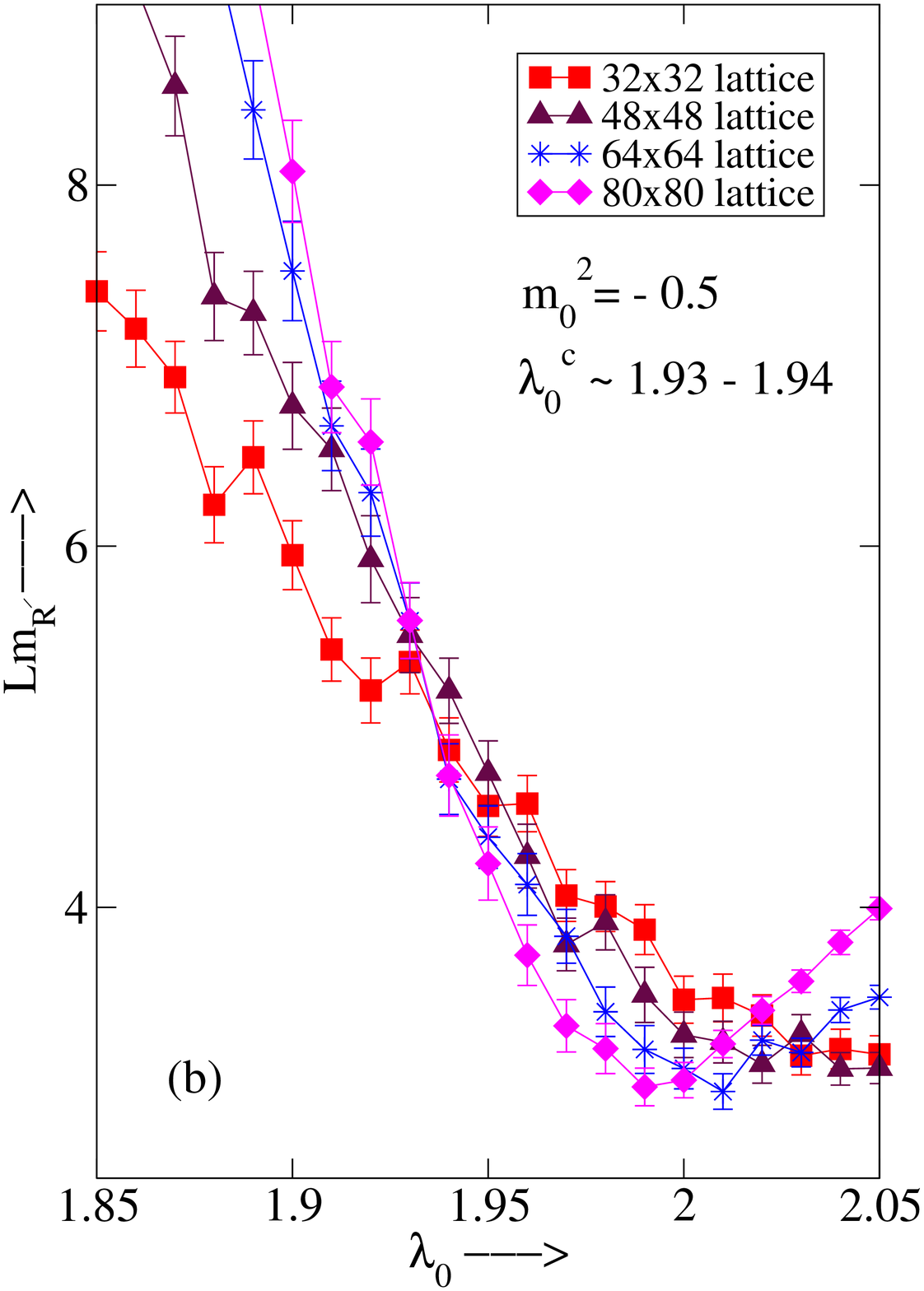}}
\caption{Determination of the critical exponents and critical coupling from
finite size scaling analysis of data for $m_R$ (or $m_{R^\prime}$) in the
critical region. $L~m_R$ and $L~m_{R^\prime}$ are plotted for different $L$
versus $\lambda_0$ with (a) $m_R$ extracted from coordinate space propagator
data and (b) $m_{R^\prime}$ extracted from a 3-point fit near zero momentum of
momentum space propagator data}
\label{m-fss}
\end{figure}     

Thus we have 
\begin{eqnarray*}
L^{\rho/\nu}/P_L(\tau) =
A_P^{-1}~A_{\xi}^{\rho/\nu}~F(A_{\xi}^{-1/\nu}~
\tau ~L^{1/\nu})
\end{eqnarray*}
where the function $F$ is the inverse of the scaling function $g$ and
$ F(A_{\xi}^{-1/\nu}~\tau ~L^{1/\nu}) \rightarrow {\rm a ~~
  constant} $ as $ \tau \rightarrow 0$ for finite $L$.
So, we can write
\be L^{\rho/\nu}/P_L(\tau) = ~A_P^{-1}~A_{\xi}^{\rho/\nu}~\left[{\rm C_P}~ +
~ {\rm D_P}~A_{\xi}^{-1/\nu}~
\tau ~ L^{1/\nu}~+ ~ {\cal O}(\tau^2)\right] \label{fss2}
\ee 
as $ \tau \rightarrow 0 $ (${\rm C_P}$ and ${\rm D_P}$  are universal
constants in the same sense as the finite size scaling functions). The
utility of Eq. (\ref{fss2}) is that if we plot 
$L^{\rho/\nu}/P_L\left(\tau\right)$ versus the coupling
$\lambda_0$ for different values of $L$, all the curves will pass through the
same point when $\tau=0$ or equivalently $ \lambda_0 = \lambda_0^c $
\cite{goldenfeld}. These ideas provide us with a very good 
method for evaluating the critical point and
checking the critical exponents.

\begin{figure}[tp]
\centering  
\includegraphics[width=3in,clip]{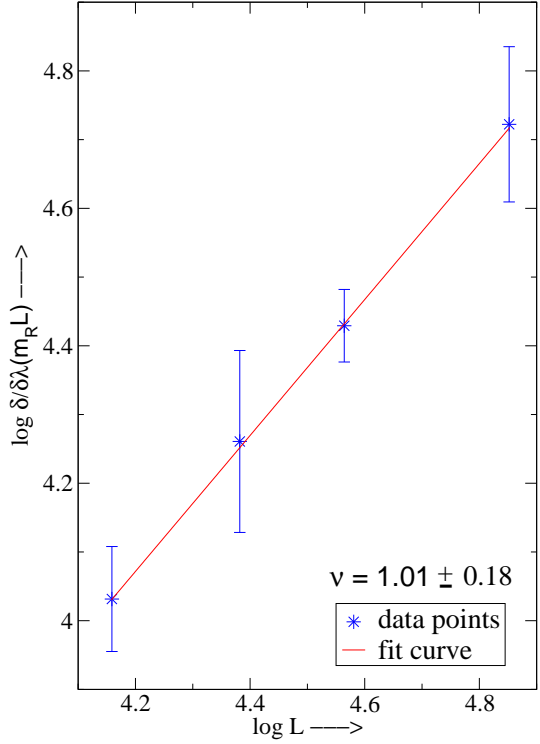}
\caption{Determination of critical exponent ($\nu$) for mass gap obtained from
  coordinate space propagator for $m_0^2 = -0.5$.}
\label{nufit}
\end{figure}

We have performed finite size scaling analysis for the observables
$~{\langle \phi\rangle}^{-1},~{m_R}^{-1} ~({\rm or} ~~{m_{R^\prime}}^{-1})$
and ${Z}^{-1}~({\rm or}~~{Z^\prime}^{-1})$.
The critical behavior of $\langle \phi
\rangle, ~~m_R$ and $Z$ may be written as 
\be
\langle \phi \rangle & = & A_{\phi}^{-1}~\tau^\beta \nonumber \\
m_R & = & A_{\xi}^{-1}~\tau^\nu \nonumber \\
Z & = & A_Z^{-1}~\tau^\eta~. \label{scaling}
\ee 
From the general expectation that in $1+1$ dimensions, 
$\phi^4$ theory and Ising
model belong to the same universality class, 
we have used the Ising values for
the corresponding exponents as inputs in our FSS analysis. Thus, $\beta =
0.125,~\nu = 1$ and $\eta = 0.25$. 

\begin{figure}[tp]
\centering  
\subfigure{
\includegraphics[width=3in,clip]{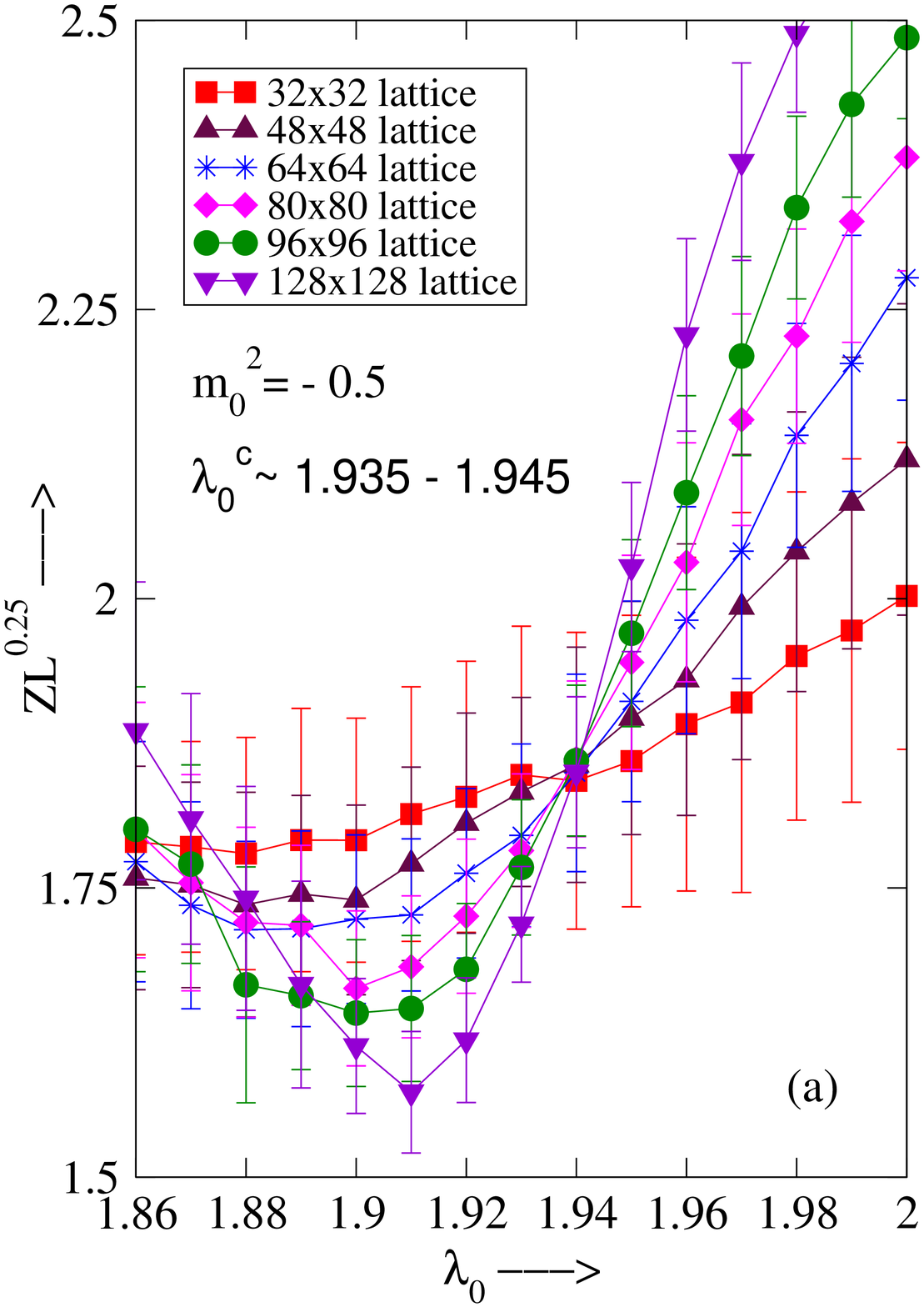}}
\subfigure{       
\includegraphics[width=3in,clip]{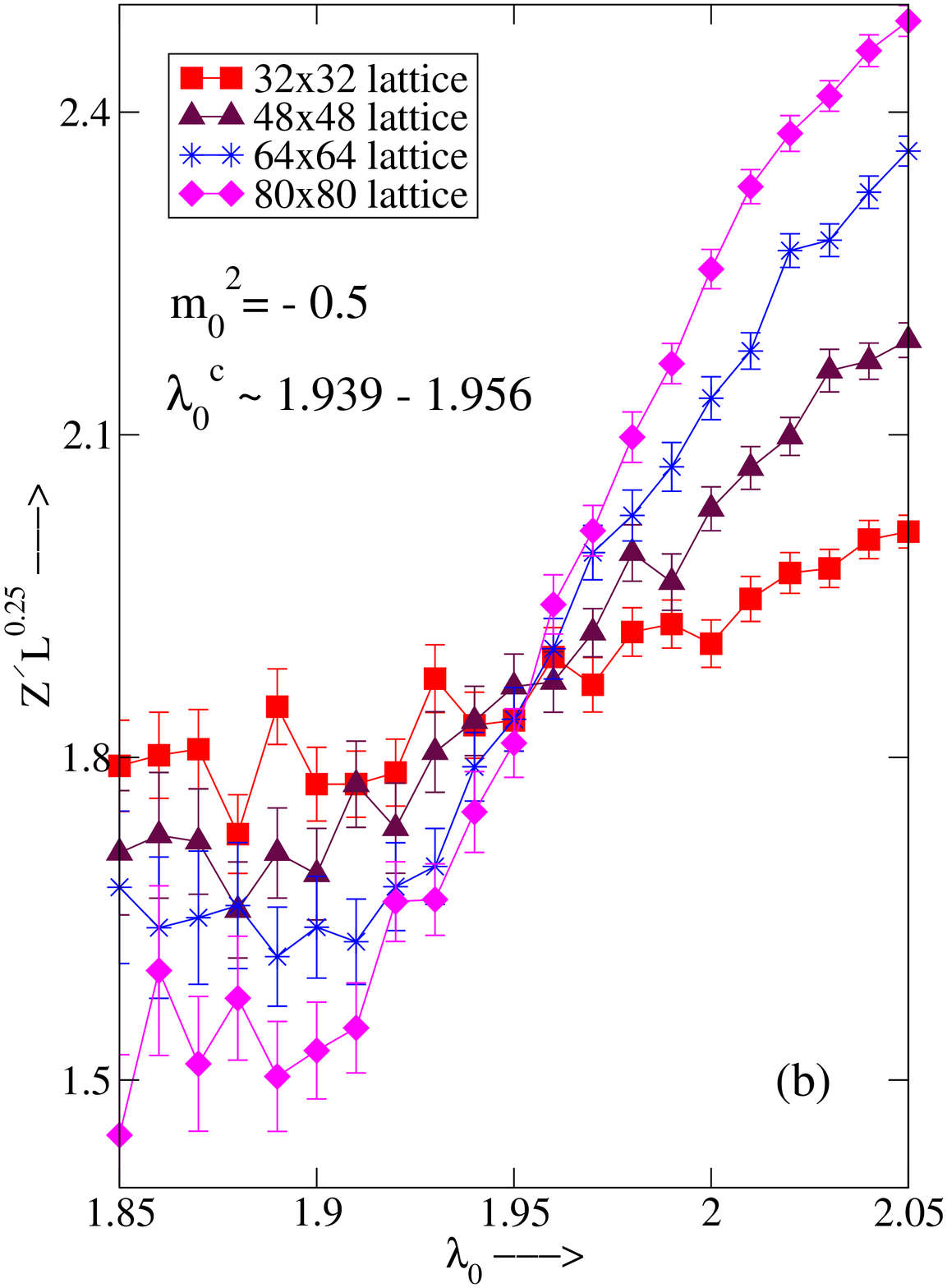}}
\caption{Determination of the critical exponents and critical coupling from
finite size scaling analysis of data for $Z$ (or $Z^\prime$) in the critical
region. $Z~L^{0.25}$ and $Z^\prime~L^{0.25}$ are plotted for different $L$
versus $\lambda_0$ with (a) $Z$ extracted from coordinate space propagator and
(b) $Z^\prime$ extracted from a 3-point fit near zero momentum of the momentum
space propagator data.}
\label{Z-fss}
\end{figure} 

According to the discussion following Eq. (\ref{fss2}), we have plotted 
$L^{\rho/\nu}/P_L(\tau)$ as a function of $\lambda_0$ near $\lambda_0^c$
for different $L$, with $~P_L(\tau)={m_R}^{-1},~Z^{-1}$ and 
$~{\langle \phi \rangle}^{-1}$ respectively. 
Figs. \ref{m-fss}(a) and (b) show the plots of $Lm_R$ and 
$Lm_{R^\prime}$ for different values of $L$ against 
$ \lambda_0$ with $m_R$ and
$m_{R^\prime}$ obtained from coordinate and momentum space propagators
respectively. All results are with $m_0^2=-0.5$. We can clearly identify the
critical point with remarkable precision from these plots and this agrees well
with the value shown in Fig. \ref{pd_cont}. Since in this case $\rho = \nu$,
we cannot verify the critical exponent for $m_R$ (or
$m_{R^\prime}$) solely from these plots. However, on differentiating
\cite{goldenfeld}, Eq. (\ref{fss2}) (with $P_L(\tau) = 1/m_R$) with respect to
$\lambda_0~$ we get 
\begin{eqnarray*}
\frac{\partial}{\partial\lambda_0}\left(L~m_R\right) = BL^{1/\nu}{\rm as}
~\tau~\rightarrow~0, 
\end{eqnarray*}
where $B$ is a constant. The exponent
$\nu$ can be computed easily by taking the logarithm of the above equation:
\be
\log\frac{\partial}{\partial\lambda_0}\left(L~m_R\right) = 
\log B~+~\frac{1}{\nu}\log L.\label{fss3} 
\ee
Determination of $\nu$ using Eq.\ (\ref{fss3}) is presented in Fig.\
\ref{nufit}.\ In this calculation, we have used the lattice data 
obtained from coordinate space propagator for four 
largest lattices $\left(64^2,~80^2,~96^2,~128^2\right)$. The value of 
$\nu$ extracted from our fitting is $1.01\pm 0.18$
which is consistent with the Ising value $(\nu = 1)$ to our
numerical accuracy.
    
\begin{figure}
\begin{minipage}[t]{8cm} 
\includegraphics[width=1\textwidth]{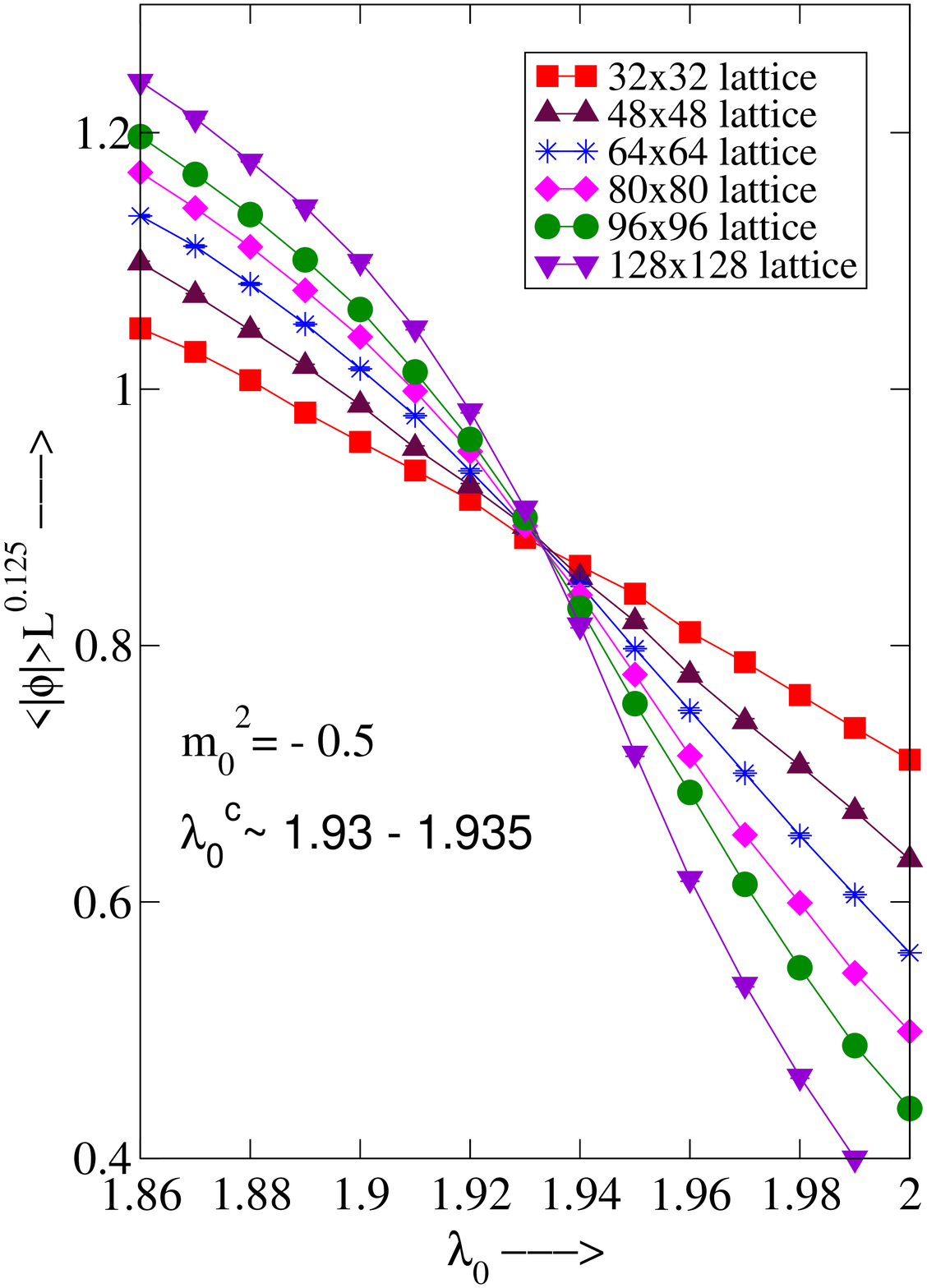}
\caption{$\langle|\phi|\rangle~L^{0.125}$  for different $L$ versus
  $\lambda_0$.} 
\label{lphi}
\end{minipage}
\hfill  
\begin{minipage}[t]{8cm}
\includegraphics[width=1\textwidth]{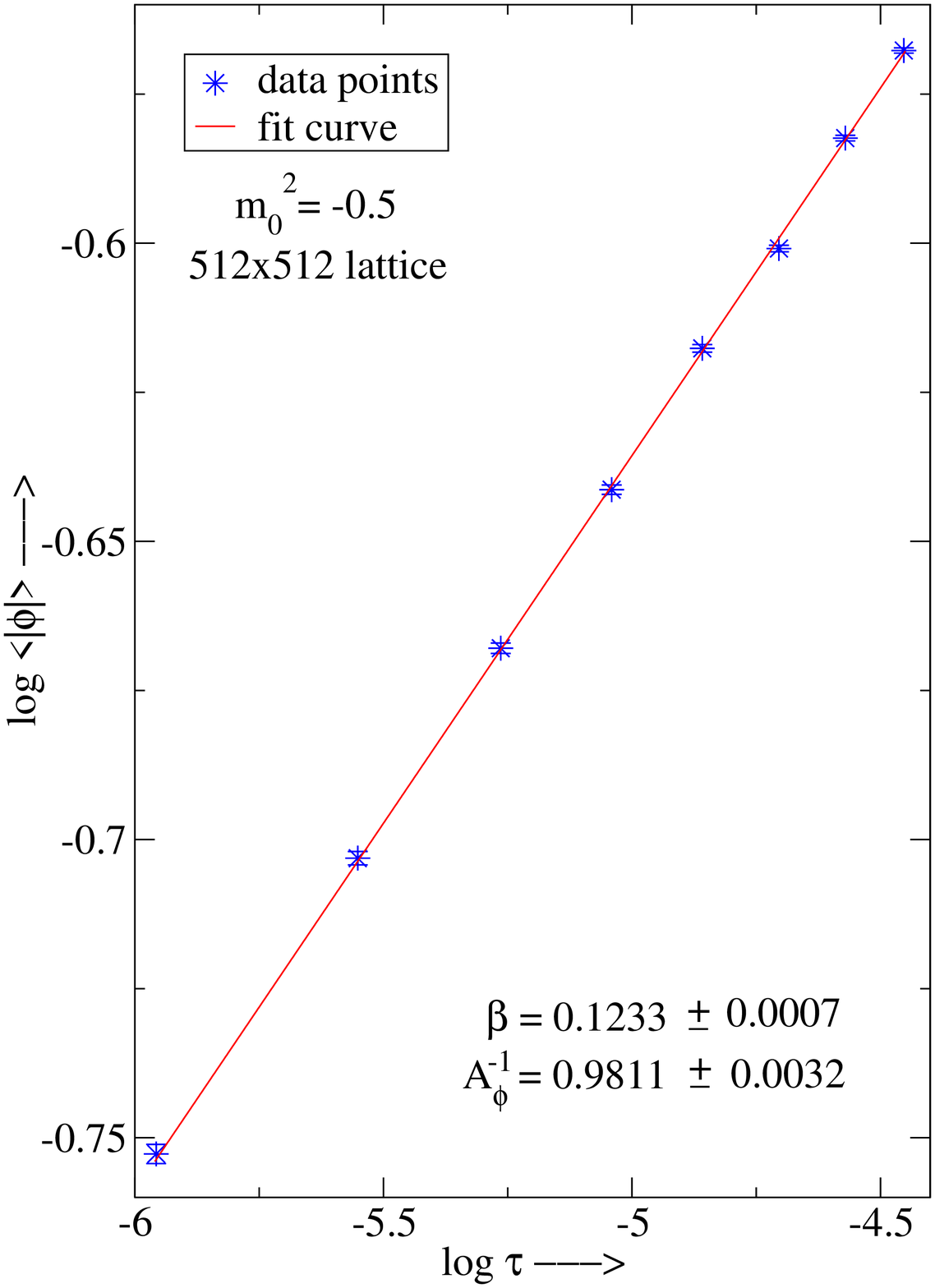}
\caption{Extraction of critical exponent associated with the order
parameter.}
\label{logfit}
\end{minipage}
\end{figure}

Fig. \ref{m-fss} (a) and (b)  also give the value of the universal
constant $C_{\xi}$ appearing in Eq. (\ref{fss2}). Applying
Eq. (\ref{fss2}) for the case of $m_R$, we find at $\tau=0$
\be
L~m_R=A_{\xi}^{-1}~A_{\xi}~C_{\xi}= C_{\xi}
\ee
because $\rho=\nu$. Discarding the $32^2$ data which seem not to
conform to FSS, from Fig. \ref{m-fss} (a) we find that $C_{\xi}\approx
5.1$. Data in Fig. \ref{m-fss} (b) is noisy; however, it still
gives a value around 5.5. 
 
Plots of $Z~L^{0.25}$ and $Z^\prime~L^{0.25}$ against $ \lambda_0$  for
lattices of different lengths $L$ with $Z$ and $Z^\prime$ computed from
coordinate and momentum space propagators are presented in
Figs. \ref{Z-fss}(a) and (b) respectively. We also present the plots of
$\langle|\phi|\rangle~L^{0.125}$ against $\lambda_0$ for different
lattice sizes in Fig. \ref{lphi}. Critical points obtained from all these FSS
plots are very close to each other. These plots also provide a very good
confirmation of the critical exponents for $Z$ and order parameter
$\langle|\phi|\rangle$. 

The critical exponent for $\langle|\phi|\rangle $ is also determined
by fitting our data for the largest lattice ($512^2$) at $m_0^2 = -0.5$ to
the corresponding scaling formula for $\langle|\phi|\rangle $. As is
shown in Fig. \ref{logfit}, the fit is very satisfactory and the
results for the critical exponent and the amplitude are:
\be
\beta=0.1233\pm 0.0007 ~~~~~{\rm and }~~~~~~ A_{\phi}^{-1}=0.9811\pm
0.0032~~~~~~(512^2~~~ {\rm lattice}) 
\ee
Also, the
critical exponent obtained from $\langle|\phi|\rangle $
data with $m_0^2=-1.0$, although not
shown here, agrees well with that extracted for $m_0^2=-0.5$. 

Within our
numerical accuracy, critical exponent for $\langle|\phi|\rangle$
calculated in the two different ways, namely, the FSS and the direct fit of
the scaling formula on the $512^2$ lattice, are very close to each other, 
indicating that $512^2$ lattice is as good as the infinite system.   

\section{Ratios $\lambda_R/m_R^2$ and $\langle\phi_R\rangle$}
We choose the following definition \cite{luwe}
of $ \lambda_{R}$, appropriate for broken phase,
in terms of the renormalized scalar mass $m_R$ (or $m_{R^\prime}$) and the
renormalized vacuum expectation value $ \langle \phi_{R} \rangle $,
\be
\lambda_{R} = 3 \frac{m_R^2 }{\langle \phi_R \rangle^2} = 
3 Z \frac{m_R^2 }{\langle \phi \rangle^2}.
\ee

\begin{figure}[tp]
\begin{centering}
\includegraphics[width=3in, clip]{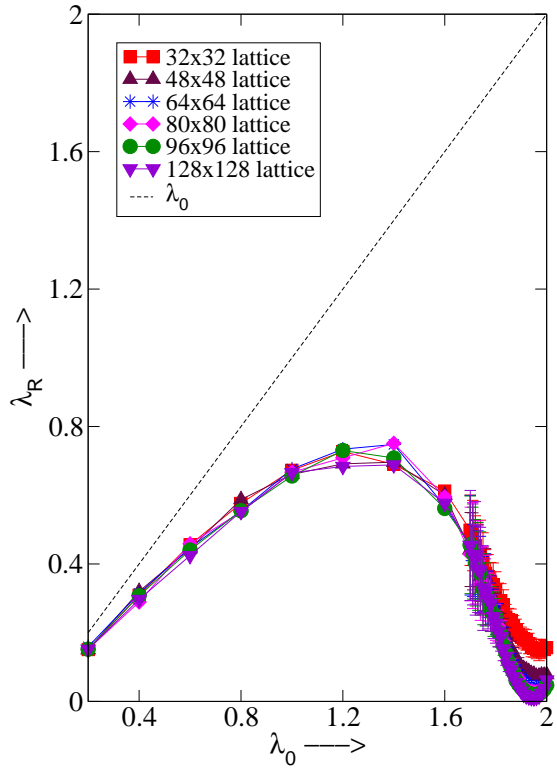}
\caption{$\lambda_R$  for different $L$ from coordinate
space propagator for $m_0^2 = -0.5$.}
\label{lr}
\end{centering}
\end{figure}

It does not require any knowledge of four point Green
function which is computationally demanding. The renormalized coupling
$\lambda_R$ calculated using $m_R$ and $Z$ from
coordinate space propagator for six different lattices using the above
method 
is presented in Fig. \ref{lr}. Error bars are not shown outside the scaling
region for the reason explained earlier. As evident from the figure,
the renormalized coupling is 
close to the tree level result in the weak coupling limit. However,
$\lambda_R$ deviates noticeably from the tree level expectation at
stronger couplings; it has a scaling behavior in the critical region
and actually vanishes at the critical point modulo finite size effects. 

In the previous section, we have already shown that the results of our
numerical analysis are consistent with the Ising values of the critical
exponents, namely, $ \beta = 0.125 $, $ \nu = 1 $ and $ \eta = 0.25 $. This
has the interesting consequence that in 1+1 dimensions, the
ratio $\lambda_R/m_R^2$ (or $\lambda_R/m_{R^\prime}^2$) is independent of the
bare couplings in the critical region, as follows:
\be
\lambda_R/{m_R}^2&=&3/{\langle \phi_R \rangle}^2~~=~~
3Z/{\langle \phi \rangle}^2 \label{lrmrph}\\
&\sim&{(\lambda_0^c-\lambda_0)}^\eta/{(\lambda_0^c-\lambda_0)}^{2\beta} 
~~=~~{(\lambda_0^c-\lambda_0)}^0, \label{uni}
\ee 
using $\eta=2\beta=0.25$.

\begin{figure}
\begin{minipage}[t]{8cm}
\includegraphics[width=1\textwidth]{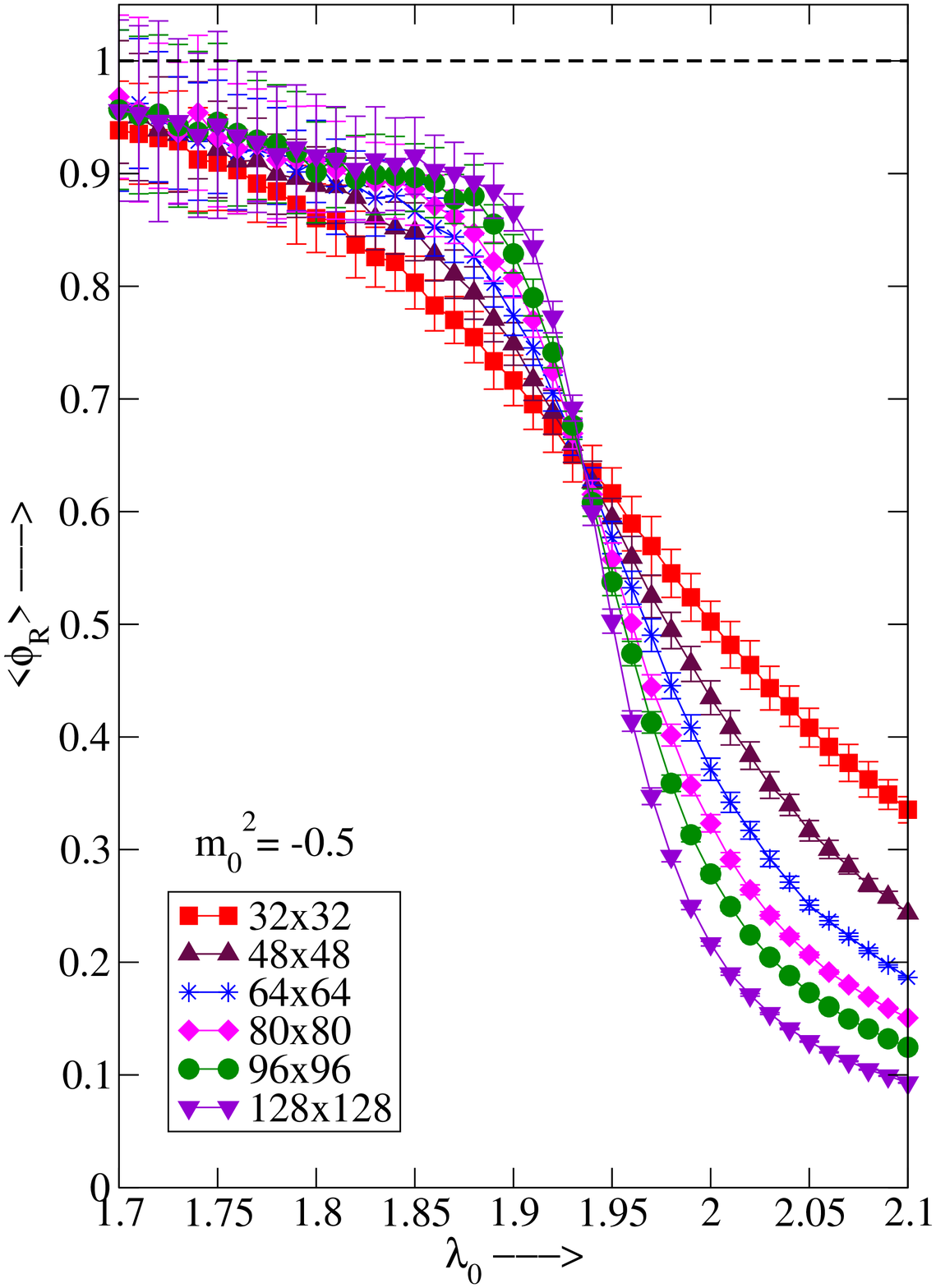}
\caption{$\langle\phi_R\rangle$ versus $\lambda_0$ for different $L$  
at\\ $m_0^2 = -0.5$.}
\label{phir1}
\end{minipage}
\hfill
\begin{minipage}[t]{8cm}
\includegraphics[width=1\textwidth]{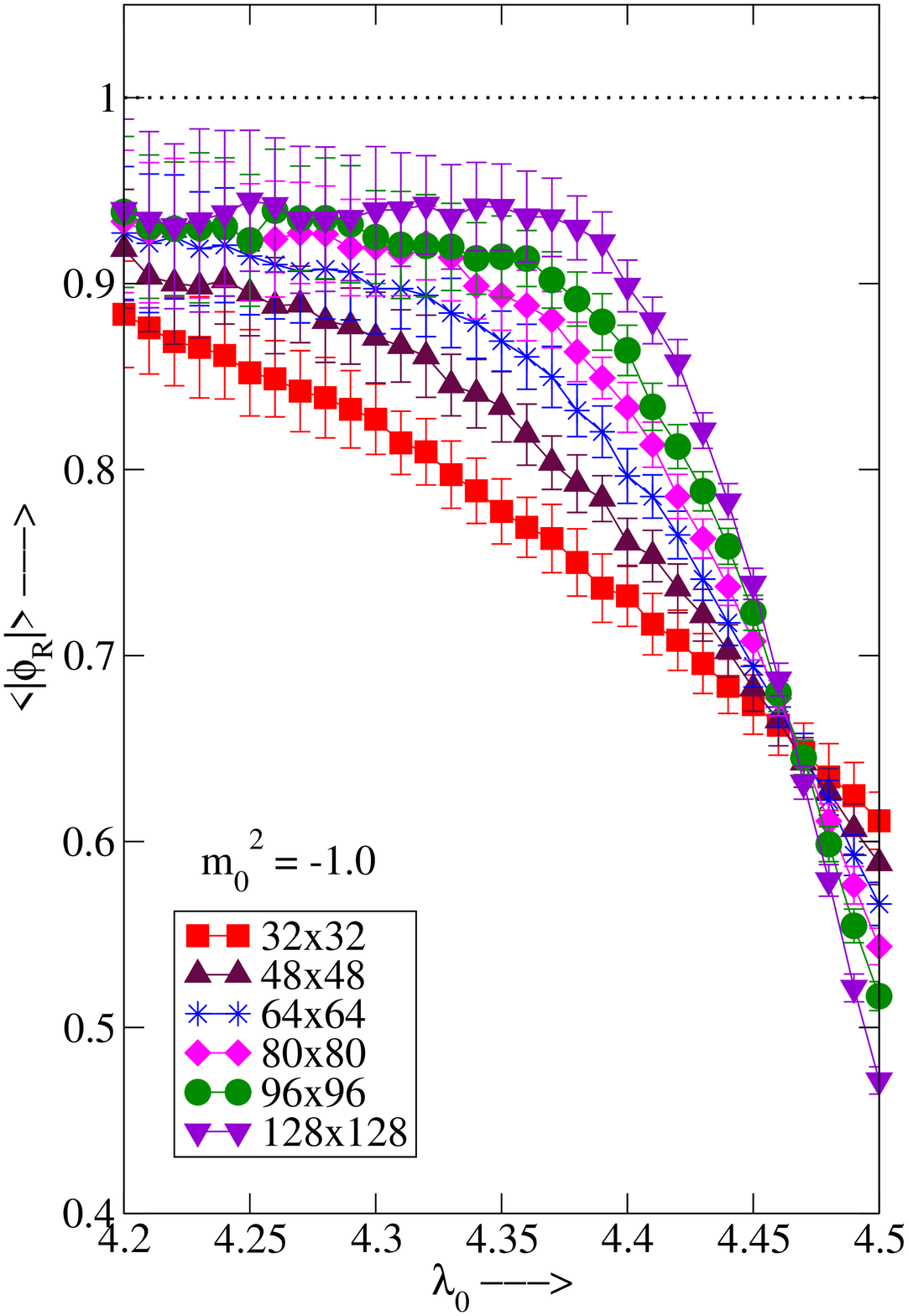}
\caption{$\langle\phi_R\rangle$  versus $\lambda_0$ for different $L$ 
at \\$m_0^2 = -1.0$.}
\label{phir2}
\end{minipage}
\end{figure}

In Figs. \ref{phir1} and \ref{phir2} for $m_0^2 = -0.5$
and $m_0^2 = -1.0$ respectively, we plot the quantity 
$\langle\phi_R\rangle$ with $Z$
evaluated from coordinate space propagator data, versus $\lambda_0$ for six
different lattice volumes for a set of bare couplings 
close to and including the critical region. 
These figures are  consistent with
Eq. (\ref{uni}) which shows that in the infinite volume limit 
$\langle\phi_R\rangle$ is independent of the bare
couplings. The figures show that for larger lattices the value of 
$\langle\phi_R\rangle$ gets close to unity along a plateau region just
away from the critical
point in the broken phase and quickly goes to a value close to zero on
the symmetric phase side. Judging from the trend in these figures, we
expect the curve in the infinite volume limit to take the shape of a
step function at the critical point with $\langle\phi_R\rangle$
dropping from around unity to zero as it passes the critical point
from the broken symmetry phase to the symmetric phase.

\begin{figure}
\begin{minipage}[t]{8cm}
\includegraphics[width=1\textwidth]{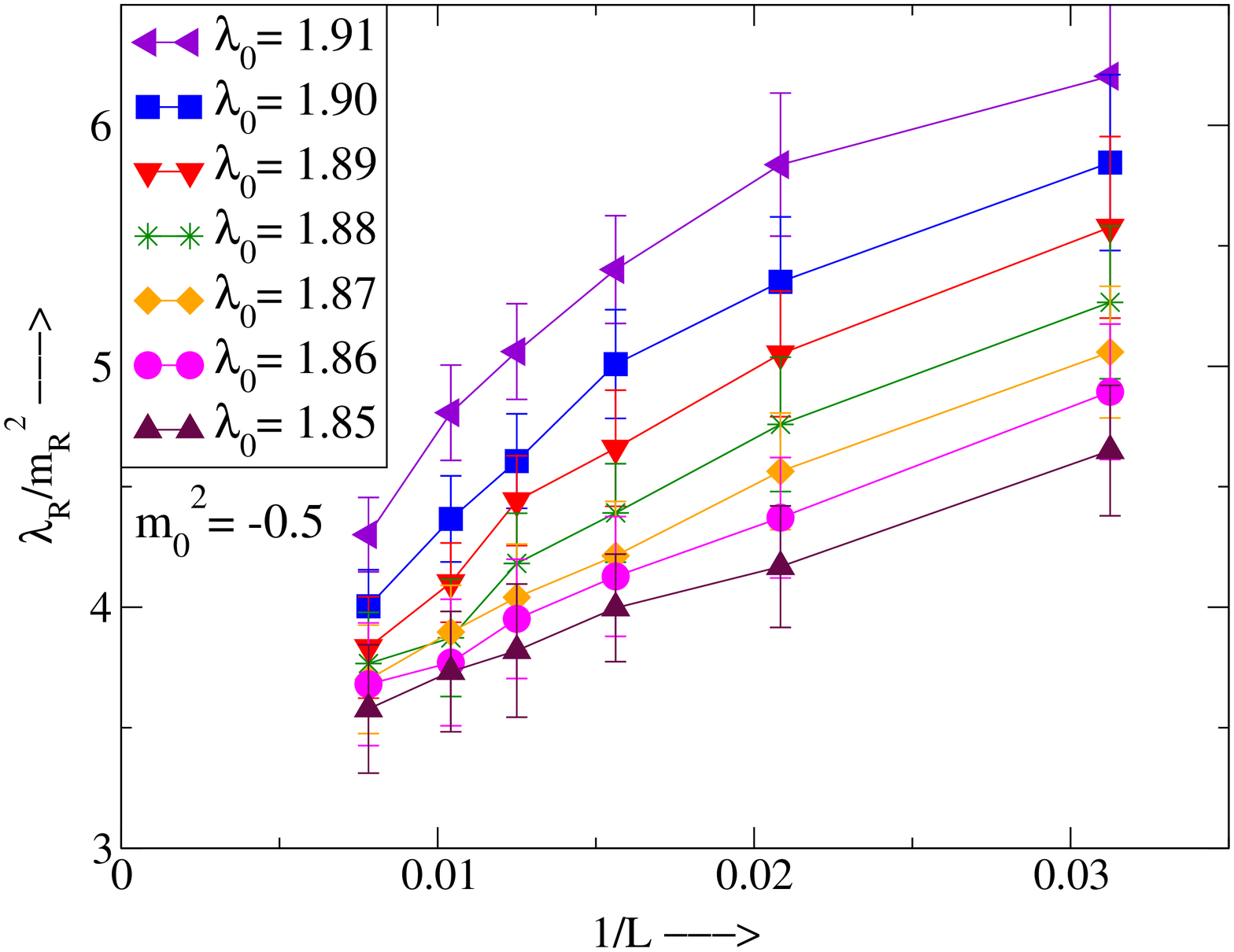}
\caption{$\lambda_R/m_R^2$ versus $1/L$ for different $\lambda_0$  
at \\ $m_0^2 = -0.5$.}
\label{ratio1}
\end{minipage}
\hfill
\begin{minipage}[t]{8cm}
\includegraphics[width=1\textwidth]{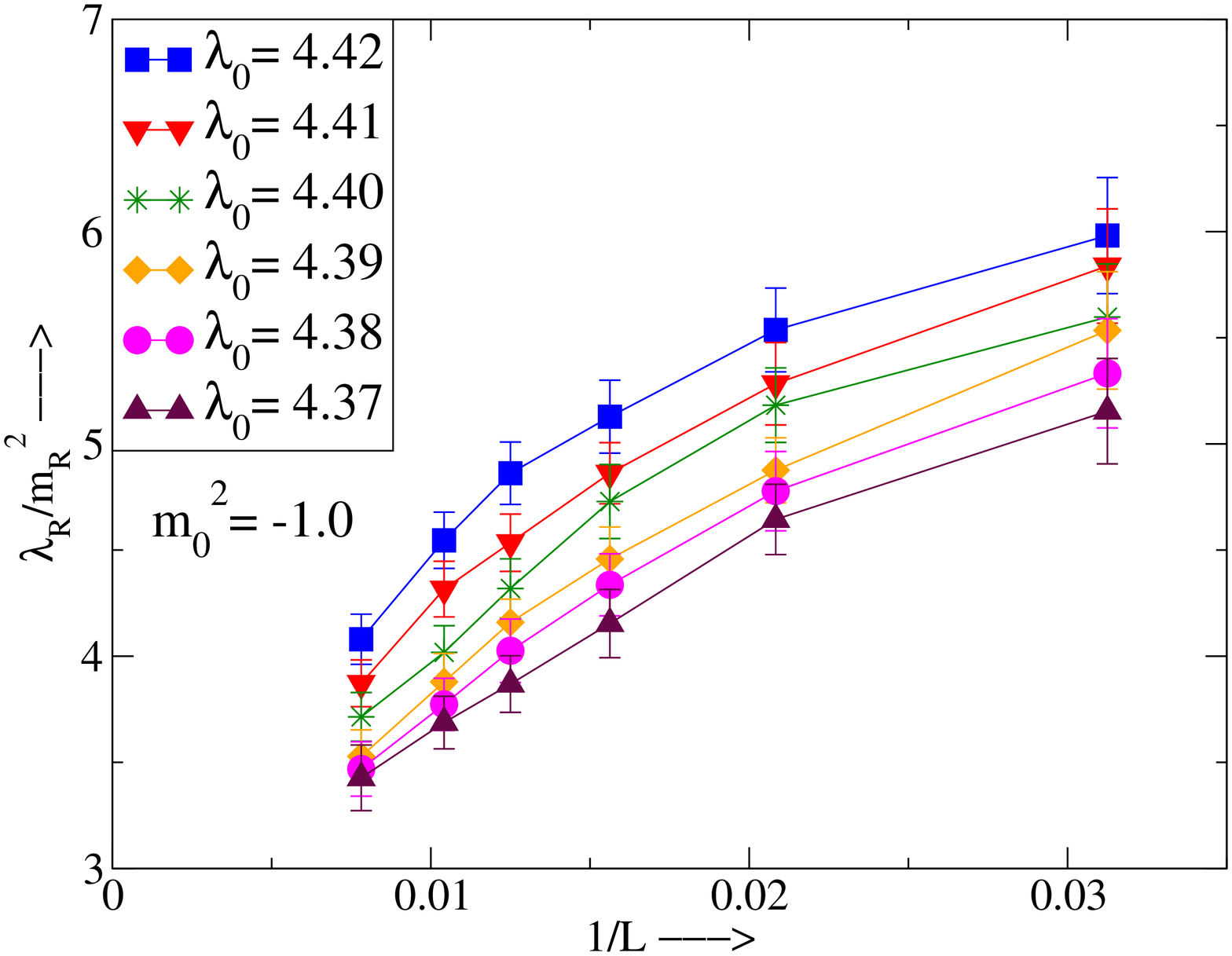}
\caption{$\lambda_R/m_R^2$  versus $1/L$ for different $\lambda_0$ 
at \\$m_0^2 = -1.0$.}
\label{ratio2}
\end{minipage}
\end{figure}

One can also try to take the infinite volume limit of 
$\langle\phi_R\rangle$ in the scaling region (as will be shown in the 
following for the ratio
$\lambda_R/m_R^2$ in Figs. \ref{ratio1} and \ref{ratio2}). From our 
extrapolations, although not shown here and already quite apparent
from Figs. \ref{phir1} and \ref{phir2}, this value seems to be very
close to unity. 

Eq. (\ref{lrmrph}) then immediately tells us that the
infinite volume limit of $\lambda_R/m_R^2$ in the scaling region would
be close to 3. This is what is indicated in Figs. \ref{ratio1} and 
\ref{ratio2}. The significant error bars in our data result mostly from
inaccuracies in the determination of the field renormalization constant
$Z$ and do not permit us to take a more accurate infinite volume
limit. However, the trend is quite unmistakable.    

We notice that for both of the two Figs. \ref{phir1} and \ref{phir2},
the curves for different volumes meet at the same value of
around 0.65 of $\langle\phi_R\rangle$ at the critical values of
$\lambda_0$ ($\sim 1.93$ for $m_0^2 = -0.5$ and $\sim 4.46$ for $m_0^2 =
-1.0$). To explain this, we need to look at Eq. (\ref{fss2}) from which
one can write, at $\tau=0$ and finite volume,
\be
\langle\phi_R\rangle=\frac{\langle\phi\rangle}{\sqrt{Z}}=\frac{\sqrt{A_Z}}
{A_{\phi}}\frac{C_{\phi}}{\sqrt{C_Z}}.   
\ee
Factors of the lattice linear dimension $L$ cancel between the
numerator and the denominator in the above equation. From the scaling
laws Eq. (\ref{scaling}) we find that the ratio
${\sqrt{A_Z}}/{A_{\phi}}$ is the
infinite volume limit of $\langle\phi_R\rangle$ which we find to be
constant irrespective of the parameters of the theory. In addition, because
the coefficients $C_{\phi}$ and $C_Z$ are constants (they may,
however, depend on things like the boundary conditions etc.), the value of 
$\langle\phi_R\rangle$ at $\tau=0$ is also constant irrespective of
the parameters and volume, as shown in
Figs. \ref{phir1} and \ref{phir2} for two sets of bare couplings. 

\section{Concluding Discussion}
Our investigation of the 1+1 dimensional $\phi^4$ theory on the lattice has
certainly turned out to be more challenging and absorbing than what
one would generally expect for a lower dimensional theory.

On the algorithmic front, since the Metropolis algorithm was extremely
inefficient near the critical region and for our study we required to 
obtain a large number of uncorrelated configurations, we had to incorporate
the cluster algorithm. Cluster algorithms are generally applicable only to
Ising-type systems. In our case, we used a result due to Wolff \cite{wolff2}
to apply the cluster algorithm to the embedded Ising variables in the $\phi^4$
theory. To update the radial modes of the fields the standard Metropolis
algorithm had to be blended with the cluster algorithm.

We explored the phase diagram of the lattice theory in two different
parameterizations. We have found that symmetry breaking occurs only with a
negative mass-squared term in the Hamiltonian. 

We needed a large number of configurations to get numerically stable results
for the connected propagators in the broken phase. Away from the critical
point, the
magnitude of the $\phi$ field is large, and the connected propagator which
would be a relatively small number had to be extracted from the subtraction
of two large numbers. In addition, the momentum space propagators showed
signs of curvature for small lattice momenta, a fact which made the
determination of the renormalized mass and the field renormalization
constant a tough one and we had to be as close to the zero momentum as
possible. 

Using a definition appropriate for broken symmetry phase, we have calculated
the renormalized coupling $\lambda_R$ which invloves $m_R,~~Z$ and
$\langle\phi\rangle$. At weak coupling limit our result is close to the tree
level result but deviates significantly in the strong coupling regime.  

We have used the finite size scaling analysis to determine the critical point
and verify and ultimately determine the critical exponents associated with
$m_R,~Z$ and $\langle\phi\rangle$. Verification of critical exponents for
$m_R$ and $Z$ are performed using the data for both coordinate and momentum
space propagators. Apart from verifying the critical exponent for
$\langle\phi\rangle$ using FSS analysis, we have also independently determined
this quantity by fitting our data for a large enough lattice
($512^2$). Our results
are consistent with the expectation that in 1+1 dimensions the $\phi^4$ theory
and the Ising model are in the same universality class.

One of the most important observation in 1+1 dimensions is that the field
renormalization constant scales with a particular critical exponent,
something that does not happen in 3+1 dimensions. This has the interesting
consequence that the renormalized field does not scale and in the infinite
volume limit, it drops from a value approximately around unity to zero
abruptly as we pass from the broken symmetry phase to the symmetric phase.  
Moreover, the ratio of the renormalized quartic
coupling to the square of the renormalized mass also does not scale and
appears to be independent of the bare parameters in the scaling
region. Numerically
this ratio seems to approach a value around 3 in the infinite volume limit. 
However, our infinite volume extrapolations are approximate due to
large finite size effects and
systematic error in the calculation of the field renormalization constant $Z$. 

For reliable extrapolation of the above
amplitude ratios to infinite volume there exist methods \cite{infvol}
which we have not tried in the present work. We need to have
smaller error bars especially on the $m_R$ and $Z$ data and this can
be taken up in a future work. 

\acknowledgments
Numerical calculations presented in this work are carried 
out  on a Power4-based IBM cluster and a Cray XD1. The High
Performance Computing Facility is supported by 
the 10$^{th}$ Five Year Plan Projects of the Theory Division, Saha
Institute of Nuclear Physics, under the DAE, Govt. of India. 


\end{document}